# Goldstone-mediated polar instability in hexagonal barium titanate

S. Simpson[1*], U. Dey[2,3], R. J. Sjökvist[4], J. Wright[5], C. Ritter[6], R. Beanland[4], N. C. Bristowe[2], and M. S. Senn[1*]

[1]*Department of Chemistry, University of Warwick, Gibbet Hill, CV4 7AL, Coventry, U.K.*

[2]*Centre for Materials Physics, Durham University, South Road, Durham, DH1 3LE, U.K.*

[3]*Luxembourg Institute of Science and Technology, Avenue des Hauts-Fourneaux 5, L4362, Esch-sur-Alzette, Luxembourg*

[4]*Department of Physics, University of Warwick, Gibbet Hill, CV4 7AL, Coventry, U.K.*

[5]*European Synchrotron Radiation Facility, BP 220 38043 Grenoble, Cedex 9, France*

[6]*Institut Laue-Langevin, 71 Avenue des Martyrs, BP 156, F-38042 Grenoble, Cedex 9, France.*

\* Email: struan.simpson@warwick.ac.uk, m.senn@warwick.ac.uk

ABSTRACT: We discover a rare structural manifestation of the Goldstone paradigm in a hexagonal polytype of the archetypal ferroelectric $BaTiO_3$. First-principles calculations confirm the Goldstone character of the order parameter, and high-resolution diffraction measurements link this to a quasi-continuous domain texture in the vicinity of the low-temperature phase transitions. Our findings highlight how changes in structural topology may be exploited to realize rich polar topologies in bulk ferroelectric perovskites.

1. Introduction

Recent years have seen an influx of interest in ferroelectric (FE) materials due to the discovery of rich topological defects such as charged domain walls [1–3], polar skyrmions [4,5], and intricate vortex patterns [6–8]. Coupling between polarization and the crystal lattice, in conjunction with crystal anisotropy, were long considered to prohibit nontrivial topologies from forming in FE materials, but recent advancements in epitaxial thin-film growth and real-space imaging techniques have since facilitated their observation [9,10]. This has prompted wider reconsideration of the capacity of FE materials to exhibit unusual domain topologies and to assess their prospects for next-generation device applications.

In general, nontrivial topologies can be stabilized when the order parameter transforms in a continuous manner. In FE nanostructures, surface-depolarizing fields enable a continuous rotation of the polarization direction at the surfaces [11,12], providing the necessary conditions for intricate topological structures to form. While this approach is infeasible in bulk FEs, extrinsic effects such as compositional disorder may also unlock continuous polarization rotation pathways, as has been shown through the observation of polar bubble domains in bulk $Bi_{0.5}Na_{0.5}TiO_3$ FEs [13]. However, such extrinsic effects are difficult to control in a reliable or dynamic fashion, hence it would be beneficial to develop alternate strategies to find polar topologies in the bulk.

Here, we propose a new mechanism to stabilize polar topological defects in bulk perovskite FEs by modifying their structural topology. Using group-theoretical methods, high-resolution diffraction measurements, and first-principles calculations, we show how transferring the second-order Jahn-Teller (SOJT) instability from the archetypal FE perovskite $BaTiO_3$ (Fig. 1a) to its six-layer hexagonal polytype (6H-$BaTiO_3$) (Fig. 1b) induces a quasi-continuous symmetry-breaking driven by the newfound Goldstone character of the order parameter. The change in structural topology results in enhanced crystal anisotropy that effectively confines the salient dipole interactions governing the wider domain texture to 2D planes (Fig. 1c), thus conferring the order parameter with continuous symmetry without requiring extrinsic compositional disorder. Our results identify a generalizable mechanism with which to attain Goldstone-mediated polar instabilities in bulk perovskite FEs, hence 6H-$BaTiO_3$ represents a model system in which to elucidate the role of structural topology and crystal anisotropy in their stabilization.

2. Methods

Pristine polycrystalline samples of 6H-$BaTiO_3$ were prepared based on a high-temperature treatment of the 3C phase [14]. Laboratory X-ray diffraction (XRD) measurements confirmed the samples typically contained only trace quantities (<0.6 wt.%) of the 3C polytype. High-resolution synchrotron XRD (S-XRD) experiments were performed at the ID22 beamline (ESRF, France), while high-resolution neutron powder diffraction (NPD) measurements were performed on the D2B beamline (ILL, France).

Symmetry-adapted Rietveld fits were performed using Topas [15,16] while additional group-theoretical analysis was performed using a combination of ISODISTORT [17,18] and INVARIANTS [19,20]. 3D-XRD data were collected at the ID11 beamline at the ESRF [21] using a 43.5 keV X-ray beam of ~120 nm in size. Measurements were performed at 150 K on an individual grain from our polycrystalline sample with approximate dimensions ~12 x 20 x 3 μm. Sinogram scans of 2D image data were collected while rotating the crystal by 180° with a 0.05° step size and translating with a step size of 50 nm over a 34 μm range. First-principles calculations were carried out within the framework of density functional theory (DFT) as implemented in the VASP code, version 6.3.2 [22,23]. Further details of the synthesis, diffraction experiments, symmetry analysis, and DFT calculations are given in the Supplementary Material.

3. Results and discussion

We begin by revisiting the crystal structure of 6H-BaTiO$_3$. The aristotype structure (Fig. 1b) is described by $P6_3/mmc$ symmetry [24], but two dielectric features on cooling [25] were concluded to reflect phase transitions to an non-polar $C222_1$ structure at $T_o \approx 220$ K and a polar $P2_1$ structure at $T_c \approx 70$ K [26,27]. Group-theoretical methods have previously shown the order parameter transforms as the $\Gamma_5^-$ irreducible representation (irrep) with respect to the $P6_3/mmc$ aristotype structure [28]. This irrep encompasses an antipolar configuration of Ti displacements (Fig. 1c). Note here we restrict our discussion to the Ti displacements as these contribute ~90% of the total $\Gamma_5^-$ mode amplitude, but minor displacements of the Ba and O atoms also occur (Table S2). Depending on the order parameter direction (OPD) of the $\Gamma_5^-$ irrep, three discrete subgroup symmetries are possible: $C222_1$, $Cmc2_1$, and $P2_1$. The two distinct branches of the $\Gamma_5^-$ (a,b) OPD represent Ti displacements aligned within the hexagonal basal plane along either the <100> ($C222_1$) or <120> ($Cmc2_1$) directions of the parent hexagonal cell (Fig. 1d), while the general (a,b) OPD comprises an admixture of these displacement directions. The choice of $\Gamma_5^-$ OPD also produces an improper ferroelastic strain transforming as $\Gamma_5^+$ (Fig. 1e) and a polar mode transforming as $\Gamma_2^-$ (Fig. 1f), with the OPD of the $\Gamma_5^+$ strain and the presence of the $\Gamma_2^-$ mode being contingent on the $\Gamma_5^-$ OPD (Fig. 1g). By treating the antipolar $\Gamma_5^-$ (a,b) and polar $\Gamma_2^-$ (a) mode components as order parameters of the form $(\eta_1,\eta_2)$ and $\eta_3$, respectively, we find that the polar mode is stabilized by a fourth-order invariant term in a Landau-style free energy expansion of the form:

$$F \sim 3\eta_1^2\eta_2\eta_3 - \eta_2^3\eta_3 \qquad (1).$$

In the $C222_1$ phase (i.e., $\eta_2 = 0$), the stabilization of the $\Gamma_2^-$ mode is negated hence the polar component only emerges for the other OPDs of the $\Gamma_5^-$ mode.

Our high-resolution S-XRD and NPD data (Figs. 2a and 2b) capture the same phase transitions reported previously [26,27]. However, we find that the low-temperature monoclinic distortion is suppressed below 40 K so that the system tends towards a re-entrant orthorhombic symmetry in the ground state

(Fig. 2c). Rietveld fits against high-resolution synchrotron XRD data collected at 10 K show that a $Cmc2_1$ structural model fits equally well to the previously proposed $P2_1$ model (Table S1, Figs. S1 and S2), so it is not possible to differentiate these models unambiguously by conventional Rietveld refinement alone. To circumvent this, we exploit the sensitivity of powder diffraction techniques to precise peak positions and treat the improper $\Gamma_5^+$ (a,b) strain mode as a proxy for the relative orientation of the $\Gamma_5^-$ OPD. This is possible because the $C222_1$ and $Cmc2_1$ phases are associated with discrete $\Gamma_5^+$ OPDs (Fig. 1g), hence there is a trivial mapping between the bases of the $\Gamma_5^-$ and $\Gamma_5^+$ OPDs. By treating the strain as a two-dimensional order parameter of the form $(\rho, \varphi)$, where $\rho = (a^2 + b^2)^{1/2}$ and $\varphi = \tan^{-1}(b/a)$, we find that the system exhibits a clear crossover between $C222_1$ and $Cmc2_1$ symmetries via the intermediate $P2_1$ phase in the 10–100 K temperature range (Fig. 2d). Any monoclinic distortion at 10 K is negligible within the very high resolution of our S-XRD experiments ($\Delta d/d \sim 10^{-4}$), so the system clearly tends towards $Cmc2_1$ symmetry in the ground state. Our first-principles calculations show that a geometry-relaxed $P2_1$ structure very closely resembles the relaxed $Cmc2_1$ structure (Table S4), showing the additional degrees of freedom in the $P2_1$ phase offer negligible structural stabilization.

Our diffraction data confirm that the phase transitions in 6H-BaTiO$_3$ are described by a progressive rotation of the $\Gamma_5^-$ displacements away from the $Cmc2_1$ ground state configuration. In this system, the <120>$_h$-oriented octahedral faces are tilted by ~9° out of the hexagonal basal plane (Fig. 3a). This means that enacting the SOJT distortion in an equivalent fashion to the 3C polytype – where the local dipoles orient towards the centers of the octahedral faces perpendicular to <111>$_c$ [29] – requires an additional out-of-plane $\Gamma_2^-$ displacive component. Our NPD data show that the local Ti environments in the 3C and 6H polytypes are very similar despite their contrasting structural topologies (Table S3), so the SOJT instability remains the dominant structure-determining force in this system. Therefore the mismatch in the orientation of the octahedral faces with the hexagonal crystal axes in the 6H polytype orthogonalizes the SOJT instability into two modes with either in-plane $\Gamma_5^-$ or out-of-plane $\Gamma_2^-$ displacive components. Increasing local fluctuations of the Ti displacements between adjacent octahedral faces rotate the average $\Gamma_5^-$ OPD away from the <120>$_h$ directions and towards the <100>$_h$ directions, eventually producing the non-polar $C222_1$ structure at higher temperatures. In this way, the $\Gamma_5^-$ OPD dictates the global symmetry so that the salient dipole interactions are effectively confined to 2D planes, thus this dimensional confinement represents the mechanism by which the order parameter can acquire a continuous symmetry [30]. This is in contrast to 3C-BaTiO$_3$, where the average orientations of locally <111>$_c$-oriented dipoles appear to be discrete throughout the entire phase diagram [29].

To fully establish the continuous symmetry and hence Goldstone character of the order parameter in 6H-BaTiO$_3$, we performed a series of single-point energy calculations by DFT and mapped out the $T = 0$ K potential energy surface (PES) around the $P6_3/mmc$ aristotype structure with respect to the $\Gamma_5^-$ mode (Fig. 3b). We obtain a characteristic Mexican hat potential with negligible energy barriers (<0.02

meV/f.u.) in the brim of the hat (Fig. 3c) that lie within the energy resolution of our DFT calculations (~0.06 meV/f.u.). This unambiguously demonstrates the U(1) symmetry and Goldstone character of the antipolar $\Gamma_5^-$ displacements in 6H-BaTiO$_3$. Our ability to resolve long-range ordered domains below $T_o$ by diffraction shows that additional couplings (i.e., either to the strain or polarization) must induce some minima in the PES to determine what the global crystallographic symmetry should be. By comparing the mode amplitudes and energy lowering from our fully relaxed and fixed-cell (zero strain) calculations, we conclude that the $\Gamma_5^+$ strain plays only a tertiary role so that coupling of the $\Gamma_5^-$ and $\Gamma_2^-$ modes of the form given in Eq. (1) introduces discrete energy minima into the PES. This conclusion is further supported by the symmetry-mode decomposition of our geometry-relaxed $Cmc2_1$ and $P2_1$ structures, which shows the amplitudes of the $\Gamma_5^+$ strain and displacive components decrease compared to the $C222_1$ structure (Table S4).

The Goldstone character of the $\Gamma_5^-$ mode should enable unusual polar topologies to emerge in this system. Similar Goldstone-like modes underpin the intricate vortex patterns which have been stabilized in improper FE hexagonal manganites of the form $R$MnO$_3$ [31,32], and they most likely drive similar behavior in CsNbW$_2$O$_9$ [33], but these systems are characterized by long-range ordered domain structures. In contrast, we have been unable to observe any long-range domain structure in 6H-BaTiO$_3$ by electron diffraction and transmission electron microscopy (TEM) (Fig. S5). Nevertheless, our ability to distinguish sharp Bragg reflections in our high-resolution S-XRD measurements demonstrates that domains must arise across length scales which we are insensitive to in our TEM experiments. For this reason, we collected 3D-XRD data on an individual grain of our polycrystalline sample and mapped out the variation in the $\Gamma_5^+$ strain across the crystallite at 150 K (note the high mechanical stability required to perform tomographic reconstructions of crystals at ~100 nm length scales currently precludes measurements below this temperature). The reconstructed strain map is shown in Fig. 4a. By parameterizing the strain in terms of $\cos(3\varphi(\Gamma_5^+))$, we account for the threefold degeneracy of the $\Gamma_5^+$ order parameter space so that regions with $C222_1$ ($\cos(3\varphi) = -1$), $Cmc2_1$ ($\cos(3\varphi) = 1$), and $P2_1$ ($-1 < \cos(3\varphi) < 1$) symmetry can be distinguished from each other. Further details of the strain reconstruction have been provided in the Supplementary Material.

Our strain map clearly shows that a highly complex microstructure is present at 150 K. The sample is predominantly described by $C222_1$ symmetry, but we find that sizeable $Cmc2_1$ and $P2_1$ inclusions on the order of ~1 μm in width prevail well above $T_c$, while curved domain boundaries indicate the low-energy nature of the intermediary domain walls. The presence of these lower-symmetry polar inclusions shows that spatially extended continuous rotation pathways for the $\Gamma_5^-$ OPD separate long-range ordered $C222_1$ domains, even above the canonical ferroelectric transition. The current temperature limitations of the 3D-XRD technique prevent us from using this method to probe the domain structure at lower temperatures, but our high-resolution S-XRD data corroborate this interpretation: multiple diffuse scattering contributions, akin to domain-wall-like scattering, arise between split diffraction peaks below $T_o$ (Fig. S10a) and persist down to 10 K. These features become particularly prominent in the vicinity of $T_c$ (see Fig. 2c), suggesting there is effectively a continuous distribution of the $\Gamma_5^-$ OPD at the canonical ferroelectric transition.

To model the domain-wall-like scattering contributions, we performed a series of strain models based on a linear interpolation between $\rho(\Gamma_5^+)$ in the $C222_1$ and $Cmc2_1$ structures (Fig. S10c). Further details of the interpolation procedure are provided in the Supplementary Material. Histograms of the phase distributions obtained from our strain-interpolated models are shown in Fig. 4b. Due to the nature of powder diffraction, we cannot resolve the absolute orientations of the different domains so instead we obtain a powder-averaged distribution at each temperature. We find that the histogram obtained at 150 K is corroborated on a qualitative basis by the distribution obtained from our 3D-XRD strain map (Fig. S14). At 10 K, the mean strain distribution is largely concentrated on the high-symmetry $Cmc2_1$ directions ($\varphi(\Gamma_5^+) = 0°, 120°, 240°$). However, the distribution becomes increasingly broadened upon approach to $T_c$ with estimated full-widths at half-maxima of ~10° and ~25° at 10 K and 40 K, respectively (Table S6, Fig. S13). Near $T_c$, the distribution collapses away from the high-symmetry directions so that a structural continuum emerges between the limiting $Cmc2_1$ and $C222_1$ ($\varphi(\Gamma_5^+) = 60°, 180°, 300°$) symmetries. Near $T_c$, $\rho(\Gamma_5^+)$ is comparable for each phase component (Table S6) hence our diffraction data effectively sample a flat cut through the PES to corroborate the Goldstone character of the $\Gamma_5^-$ mode. Above $T_c$, the strain distributions gradually recondense around the $C222_1$ directions, but unusually we find that the system once again disorders between all possible orientations of the $\Gamma_5^-$ OPD upon approach to $T_o$. This is consistent with the restoration of the emergent U(1)-like symmetry of the order parameter for small $\rho(\Gamma_5^-)$ in the vicinity of $T_o$, hence further validating the Goldstone character of the $\Gamma_5^-$ mode.

Our discovery of a Goldstone-like mode in 6H-BaTiO$_3$ suggests analogous polarization rotation pathways could arise in other perovskite polytypes. This was also indicated by the recent theoretical prediction of a Goldstone mode in 2H-BaMnO$_3$ [34]. We have further explored this possibility based on DFT-relaxed structures of simulated polytypes of BaTiO$_3$ (Fig. S15, Tables S7–10). Our phonon calculations show that the $\Gamma_5^-$ mode is unstable for the 4H, 8H, and 10H polytypes, with the $C222_1$, $Cmc2_1$, and $P2_1$ structures each possessing very similar energies. While we have not explicitly computed the full PES for these polytypes, the energetic similarity of their low-symmetry phases implies that the Goldstone character of the SOJT instability is retained despite them representing distinct transformations of the basic 3C perovskite topology. Goldstone-like modes hence appear to be generalizable across numerous perovskite polytypes, showing they form a rich compositional space in which to discover intriguing polar topological structures as well.

## 4. Conclusions

Our symmetry-adapted parametrization of the order parameter in 6H-BaTiO$_3$ has allowed us to explicitly link its Goldstone-like mode to a quasi-continuous domain texture in the vicinity of its low-temperature phase transitions. 6H-BaTiO$_3$ represents a model system in which to unravel the influence of structural topology and crystal anisotropy in stabilizing unusual domain topologies in FE perovskites, with the eventual goal of understanding how these topologies can be controlled in device applications. Coupling of the ferroelastic strain to the order parameter in these polytypes presents intriguing opportunities to further tune the domain texture through strain engineering, and this comprises a promising route by which to stabilize further exotic topological textures in either the bulk or thin-film limits. The apparent generality of our results to other perovskite polytypes highlights further opportunities to stabilize Goldstone-like modes or rich polar topologies in bulk FE perovskites, prompting deeper consideration of the exotic microstructural phenomena which have yet to be encountered in these systems.

Acknowledgements

We acknowledge the European Synchrotron Radiation Facility (ESRF) for the provision of synchrotron facilities under proposal numbers MA5867 (doi:10.15151/ESRF-ES-1302904542) and HC5590 (doi:10.15151/ESRF-ES-1734554406). We thank the Institut Laue-Langevin for providing neutron facilities under experiment number 5-24-716 (doi:10.5291/ILL-DATA.5-24-71). We thank A. N. Fitch and C. Dejoie for support in using ID22 at the ESRF. Initial sample characterization was performed at the Warwick X-ray Research Technology Platform as well as Diamond Light Source under the Oxford-Warwick Solid-State Chemistry BAG (CY39378). S.S. and M.S.S acknowledge funding from the Royal Society (UF160265 and URF\R\231012). U.D. and N.C.B. acknowledge the Leverhulme Trust for a research project grant (Grant No. RPG-2020-206) and the Hamilton HPC Service of Durham University.

References

[1] D. Meier, J. Seidel, A. Cano, K. Delaney, Y. Kumagai, M. Mostovoy, N. A. Spaldin, R. Ramesh, and M. Fiebig, Nat. Mater. **11**, 284 (2012).

[2] Y. S. Oh, X. Luo, F.-T. Huang, Y. Wang, and S.-W. Cheong, Nat. Mater. **14**, 407 (2015).

[3] R. G. P. McQuaid, M. P. Campbell, R. W. Whatmore, A. Kumar, and J. M. Gregg, Nat. Commun. **8**, 15105 (2017).

[4] S. Das, Y. L. Tang, Z. Hong, M. A. P. Gonçalves, M. R. McCarter, C. Klewe, K. X. Nguyen, F. Gómez-Ortiz, P. Shafer, E. Arenholz et al., Nature **568**, 368 (2019).

[5] R. Zhu, Z. Jiang, X. Zhang, X. Zhong, C. Tan, M. Liu, Y. Sun, X. Li, R. Qi, K. Qu, et al., Phys. Rev. Lett. **129**, 107601 (2022).

[6] T. Choi, Y. Horibe, H. T. Yi, Y. J. Choi, W. Wu, and S.-W. Cheong, Nat. Mater. **9**, 253 (2010).

[7] S. C. Chae, N. Lee, Y. Horibe, M. Tanimura, S. Mori, B. Gao, S. Carr, and S.-W. Cheong, Phys. Rev. Lett. **108**, 167603 (2012).

[8] G. Sánchez-Santolino, V. Rouco, S. Puebla, H. Aramberri, V. Zamora, M. Cabero, F. A. Cuellar, C. Munuera, F. Mompean, M. Garcia-Hernandez, et al., Nature **626**, 529 (2024).

[9] J. M. Gregg, Ferroelectrics **433**, 74 (2012).

[10] J. Junquera, Y. Nahas, S. Prokhorenko, L. Bellaiche, J. Íñiguez, D. G. Schlom, L.-Q. Chen, S. Salahuddin, D. A. Muller, L. W. Martin, and R. Ramesh, Rev. Mod. Phys. **95**, 025001 (2023).

[11] H. Fu and L. Bellaiche, Phys. Rev. Lett. **91**, 257601 (2003).

[12] I. I. Naumov, L. Bellaiche, and H. Fu, Nature **432**, 737 (2004).

[13] J. Yin, H. Zong, H. Tao, X. Tao, H. Wu, Y. Zhang, L.-D. Zhao, X. Ding, J. Sun, J. Zhu, J. Wu, and S. J. Pennycook, Nat. Commun. **12**, 3632 (2021).

[14] K. W. Kirby and B. A. Wechsler, J. Am. Ceram. Soc. **74**, 1841 (1991).

[15] J. S. O. Evans, Materials Science Forum **651**, 1 (2010).

[16] B. J. Campbell, J. S. O. Evans, F. Perselli, and H. T. Stokes, IUCr Comput. Comm. Newsl. **8**, 81 (2007).

[17] B. J. Campbell, H. T. Stokes, D. E. Tanner, and D. M. Hatch, J. Appl. Cryst. **39**, 607 (2006).

[18] H. T. Stokes, D. M. Hatch, and B. J. Campbell, iso.byu.edu.

[19] H. T. Stokes, D. M. Hatch, and B. J. Campbell, iso.byu.edu.

[20] D. M. Hatch and H. T. Stokes, J. Appl. Cryst. **36**, 951 (2003).

[21] J. Wright, C. Giacobbe, and M. Majkut, Curr. Opin. Solid State Mater. Sci. **24**, 100818 (2020).

[22] G. Kresse and J. Furthmüller, Comput. Mater. Sci. **6**, 15 (1996).

[23] G. Kresse and J. Furthmüller, Phys. Rev. B **54**, 11169 (1996).

[24] R. D. Burbank and H. T. Evans, Acta Cryst. **1**, 330 (1948).

[25] E. Sawaguchi, Y. Akishige, and M. Kobayashi, Jpn. J. Appl. Phys. **24**, 252 (1985).

[26] Y. Noda, K. Akiyama, T. Shobu, Y. Kuroiwa, H. Nakao, Y. Morii, and H. Yamaguchi, Ferroelectrics **217**, 1 (1998).

[27] Y. Noda, K. Akiyama, T. Shobu, Y. Morii, N. Minakawa, and H. Yamaguchi, Journal of Physics and Chemistry of Solids **60**, 1415 (1999).

[28] D. M. Hatch and H. T. Stokes, Phys. Rev. B **40**, 4198 (1989).

[29] M. S. Senn, D. A. Keen, T. C. A. Lucas, J. A. Hriljac, and A. L. Goodwin, Phys. Rev. Lett. **116**, 207602 (2016).

[30] J. M. Kosterlitz and D. J. Thouless, J. Phys. C: Solid State Phys. **6**, 1181 (1973).

[31] S.-Z. Lin, X. Wang, Y. Kamiya, G.-W. Chern, F. Fan, D. Fan, B. Casas, Y. Liu, V. Kiryukhin, W. H. Zurek et al., Nat. Phys. **10**, 970 (2014).

[32] Q. N. Meier, A. Stucky, J. Teyssier, S. M. Griffin, D. van der Marel, and N. A. Spaldin, Phys. Rev. B **102**, 014102 (2020).

[33] J. A. McNulty, T. T. Tran, P. S. Halasyamani, S. J. McCartan, I. MacLaren, A. S. Gibbs, F. J. Y. Lim, P. W. Turner, J. M. Gregg, P. Lightfoot, and F. D. Morrison, Adv. Mater. **31**, 1903620 (2019).

[34] X. Zhang, Q.-J. Ye, and X.-Z. Li, Phys. Rev. B **103**, 024101 (2021).

[35] A. S. Gibbs, K. S. Knight, and P. Lightfoot, Phys. Rev. B **83**, 094111 (2011).

[36] D. C. Sinclair, J. M. S. Skakle, F. D. Morrison, R. I. Smith, and T. P. Beales, J. Mater. Chem. **9**, 1327 (1999).

[37] P. E. Blöchl, Phys. Rev. B **50**, 17953 (1994).

[38] G. Kresse and D. Joubert, Phys. Rev. B **59**, 1758 (1999).

[39] J. P. Perdew, A. Ruzsinszky, G. I. Csonka, O. A. Vydrov, G. E. Scuseria, L. A. Constantin, X. Zhou, and K. Burke, Phys. Rev. Lett. **100**, 136406 (2008).

[40] A. Togo and I. Tanaka, Scr. Mater. **108**, 1 (2015).

[41] X. Gonze, J.-C. Charlier, D. C. Allan, and M. P. Teter, Phys. Rev. B **50**, 13035 (1994).

[42] X. Gonze and C. Lee, Phys. Rev. B **55**, 10355 (1997).

[43] S. Baroni, S. de Gironcoli, A. Dal Corso, and P. Giannozzi, Rev. Mod. Phys. **73**, 515 (2001).

[44] M. Gajdoš, K. Hummer, G. Kresse, J. Furthmüller, and F. Bechstedt, Phys. Rev. B **73**, 045112 (2006).

[45] G. H. Kwei, A. C. Lawson, S. J. L. Billinge, and S. W. Cheong, J. Phys. Chem. **97**, 2368 (1993).

[46] ImageD11, github.com/FABLE-3DXRD/ImageD11.

[47] J. P. Wright, C. Giacobbe, and E. Lawrence Bright, Crystals **12**, (2022).

[48] A. Bonnin, J. P. Wright, R. Tucoulou, and H. Palancher, Appl. Phys. Lett. **105**, 084103 (2014).

[49] A. Henningsson and S. A. Hall, Acta Cryst. A**79**, 542 (2023).

[50] W. van Aarle, W. J. Palenstijn, J. Cant, E. Janssens, F. Bleichrodt, A. Dabravolski, J. D. Beenhouwer, K. J. Batenburg, and J. Sijbers, Opt. Express **24**, 25129 (2016).

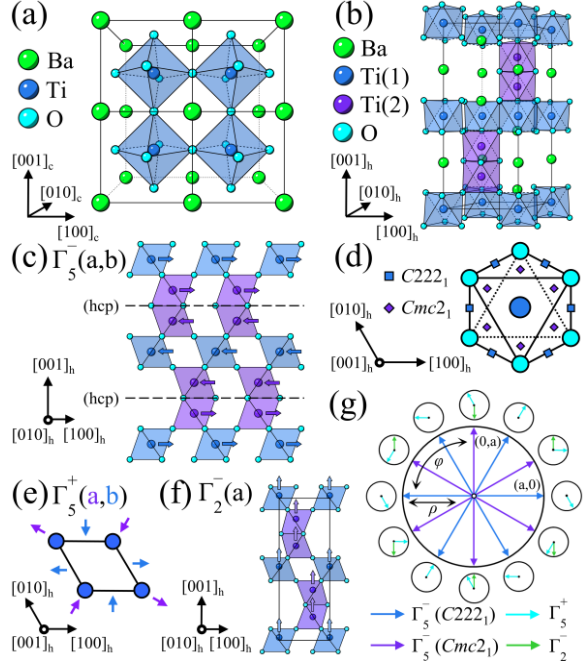

FIG. 1. (a) and (b) The aristotype crystal structures of the cubic (3C) and hexagonal (6H) polytypes, respectively. (c) The Goldstone-like mode in 6H-BaTiO$_3$. Dashed lines denote hexagonal close-packed (hcp) BaO$_3$ layers, across which the in-plane orientation of the antipolar $\Gamma_5^-$ displacements switches by 180°; the Ba atoms have been omitted for visual clarity. (d) Projection of the high-symmetry OPDs of the $\Gamma_5^-$ displacements onto an individual TiO$_6$ octahedron. (e) and (f) depict the improper ferroelastic strain ($\Gamma_5^+$) and polar ($\Gamma_2^-$) modes, respectively. (g) Relationships between the antipolar, ferroelastic, and FE domains in terms of the order parameter spaces of the $\Gamma_5^-$, $\Gamma_5^+$, and $\Gamma_2^-$ modes. The large circle depicts the $\Gamma_5^-$ order parameter space, while the outer circles show how the $\Gamma_5^-$ OPD propagates to the $\Gamma_5^+$ and $\Gamma_2^-$ modes. Variations of the amplitude ($\rho$) and phase ($\varphi$) of the $\Gamma_5^-$ order parameter have been illustrated for reference.

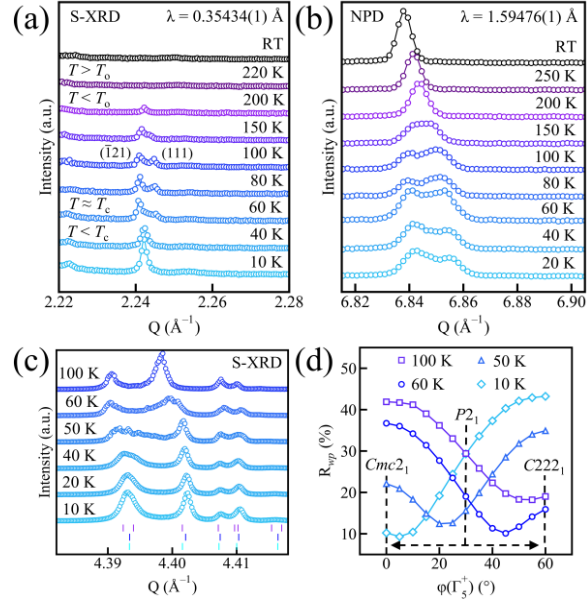

FIG. 2. Excerpts of the low-temperature S-XRD (a) and NPD (b) patterns, tracking the evolution of the $C222_1$ ($T < T_o$), $P2_1$ ($T \approx T_c$), and $Cmc2_1$ ($T < T_c$) phases. (c) Suppression of monoclinic peak splitting in the S-XRD patterns near and below $T_c$. Expected Bragg reflections for the $C222_1$ (cyan ticks), $Cmc2_1$ (blue ticks), and $P2_1$ (purple ticks) phases at 50 K are shown below the diffraction patterns for reference. (d) Fitting statistics from strain-mode refinements against S-XRD data based on fixed variations in $\varphi(\Gamma_5^+)$.

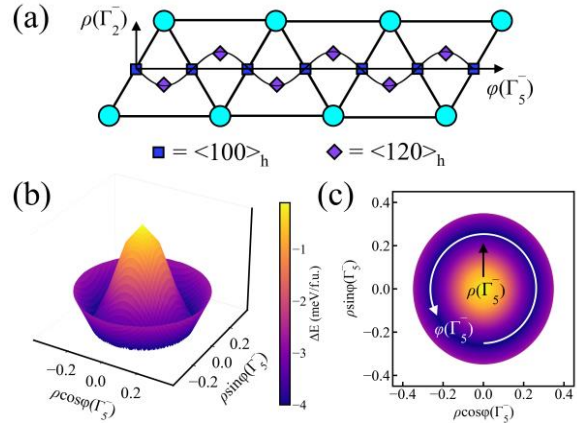

FIG. 3. (a) Schematic of a TiO$_6$ octahedron "unfolded" with respect to the hexagonal basal plane, mapping the polarization rotation pathway in terms of the OPD of the $\Gamma_5^-$ mode and the amplitude of the $\Gamma_2^-$ mode ($\varphi(\Gamma_5^-)$ and $\rho(\Gamma_2^-)$, respectively). The <100>$_h$ and <120>$_h$ directions are highlighted with blue squares and purple diamonds, respectively. (b) PES of 6H-BaTiO$_3$ with respect to the $\Gamma_5^-$ mode. (c) 2D projection of (b), with variations of $\varphi(\Gamma_5^-)$ and $\rho(\Gamma_5^-)$ depicted for reference.

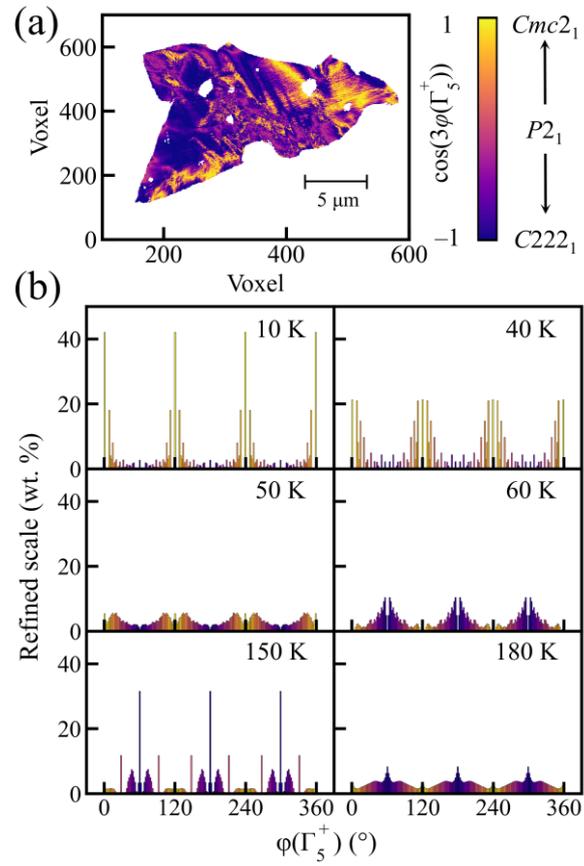

FIG. 4. (a) 3D-XRD strain map based on a least-squares refinement of the local cell parameters of an individual 6H-BaTiO$_3$ crystallite measured at 150 K. (b) Histograms depicting the powder-averaged domain distributions obtained at select temperatures based on a strain-interpolated model between the limiting $C222_1$ (purple) and $Cmc2_1$ (yellow) strain states.

Supplementary information for "Goldstone-mediated polar instability in hexagonal barium titanate".

Supplementary Note 1. Details of the synthetic conditions used to prepare 6H-BaTiO$_3$.

To prepare 6H-BaTiO$_3$, a 3C-BaTiO$_3$ precursor was synthesized by grinding together stoichiometric quantities of BaCO$_3$ and TiO$_2$ before heating the combined mixture to 900 °C for 10 h. The precursor was then pelletized and transferred to a Pt crucible, where it was subsequently annealed at 1525 °C for at least 10 h. Samples were quenched from 1525 °C into liquid nitrogen to kinetically trap the 6H phase; quenching in air was found to produce large quantities of the cubic phase (>3 wt.%). Subsequent Rietveld refinements against synchrotron XRD (S-XRD) and neutron powder diffraction (NPD) data confirmed that the Ba, Ti, and O sites could all be refined to within 1% of their nominal occupancies, showing our samples are stoichiometric within the resolution of our experiments. Additionally, the intradimer Ti(2)–Ti(2) distances in our samples (2.682(1) Å) are in good agreement with those previously reported in oxygen-stoichiometric samples (2.690(4) Å) [36], substantiating that the nominal oxygen stoichiometry of our samples is accurate. Although 6H-BaTiO$_3$ comprises a metastable polytype of the perovskite structure, we found our samples remain air-stable for up to a year with no noticeable signs of any structural degradation to the 3C phase.

For the high-resolution S-XRD measurements, a sample of 6H-BaTiO$_3$ was loaded into a 0.3 mm borosilicate capillary and a 35 keV X-ray beam was selected to minimize absorption from the heavy Ba$^{2+}$ cations. Data were collected in the temperature range 10–290 K for ~50 minutes at each temperature point. For the high-resolution NPD measurements, 3 g of 6H-BaTiO$_3$ were loaded into an 8 mm vanadium can. Data were collected in the range 20–290 K for ~2 h at each temperature. Our samples were extremely crystalline, meaning the Bragg peaks in the NPD patterns were sharper than the available instrumental standards. To create the instrumental model for our Rietveld fits, the instrumental parameters were refined against the data collected at 290 K (i.e., in the aristotype *P6$_3$/mmc* phase). These parameters were then fixed for the subsequent Rietveld fits at lower temperatures so that any additional peak broadening in the low-temperature phases could be attributed to the sample.

Supplementary Note 2. Details of the first-principles calculations performed.

Density functional theory (DFT) calculations were performed using the plane wave augmented (PAW) method implemented in the Vienna Ab initio Simulation Package (VASP), version 6.3.2 [22, 23]. PAW pseudopotentials (PBE, version 5.4) [37,38] were used with 10 valence electrons for Ba ($5s^2 5p^6 6s^2$), 10 for Ti ($3p^6 4s^2 3d^2$), and 6 for O ($2s^2 2p^4$). We employed the PBEsol general gradient approximation (GGA) [39] to the exchange correlation functional. Convergence tests performed on the aristotype $P6_3/mmc$ structure of 6H-BaTiO$_3$ revealed that a plane wave cutoff of 700 eV and a 6 × 6 × 2 k-mesh grid were sufficient to resolve the total energies, forces and stresses within 0.06 meV/formula unit, 1 meV/Å and 0.1 GPa, respectively. We set the energy convergence criterion at $10^{-9}$ eV for all the calculations and carried out full structural relaxations until the Hellmann-Feynman forces on each atom were less than 0.1 meV/Å. Phonon spectra were computed employing the finite displacement method implemented in PHONOPY [40] using 2 × 2 × 2 phonon supercells. The phonon dispersions for the $C222_1$ and $Cmc2_1$ phases were converted to the primitive basis using the transformation matrix:

$$\begin{pmatrix} 0.5 & -0.5 & 0 \\ 0.5 & 0.5 & 0 \\ 0 & 0 & 1 \end{pmatrix}$$

To account for the long-range dipole-dipole interaction near the zone center, the non-analytical term correction to the dynamical matrix was included [41,42] by computing the Born effective charges and the static dielectric tensor using VASP-DFPT calculations [43,44].

Calculated phonon dispersion curves for 6H-BaTiO$_3$ are presented in Fig. S1. We find that both of the $\Gamma_5^-$ and $\Gamma_2^-$ modes are unstable with respect to the aristotype $P6_3/mmc$ structure, which is in accordance with the experimental results. The geometry-relaxed structures are discussed in more detail in Supplementary Note 5.

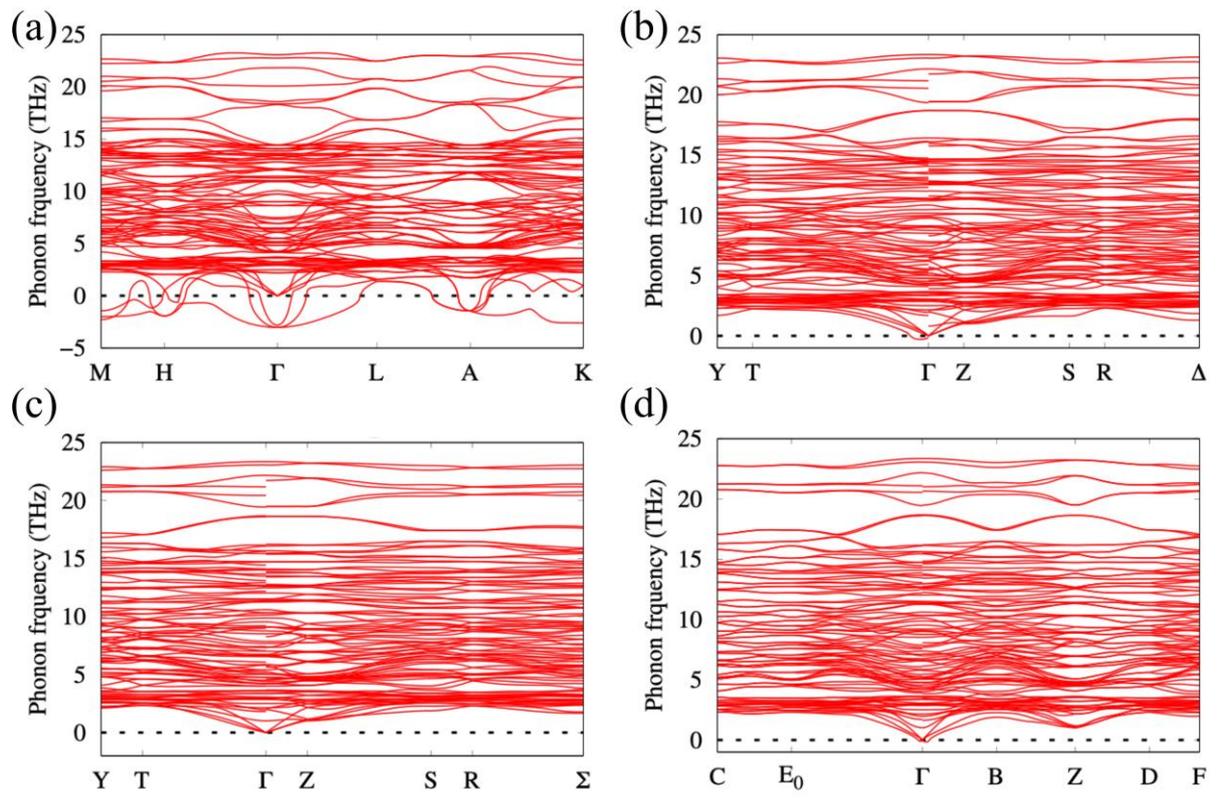

Figure S1. Phonon dispersions of the $P6_3/mmc$ (a), $C222_1$ (b), $Cmc2_1$ (c), and $P2_1$ (d) structures of 6H-BaTiO$_3$.

Supplementary Note 3. Details of symmetry-mode Rietveld refinements used to assess the appropriate low-temperature symmetry assignments in 6H-BaTiO$_3$.

Symmetry-mode Rietveld refinements were performed against the low-temperature S-XRD and NPD datasets using $C222_1$, $Cmc2_1$, and $P2_1$ structural models. These fits were performed to assess the most appropriate symmetry assignments below $T_o$ and $T_c$. A comparison of the fitting statistics obtained with these models at select temperatures is given in Table S1, while representative fits are shown in Figs. S2 and S3. Representative crystallographic information files for each model are provided as part of the Supplementary Information. The S-XRD instrumental peak shape was modelled using a simple convolution of fixed Gaussian and Lorentzian contributions obtained by refinement against a Si standard, while the NPD peak shape was modelled using a fixed pseudo-Voigt profile obtained by modelling the peak shapes of the room-temperature structure. The following parameters were refined in all models: background terms, consisting of a shifted Chebyschev polynomial (containing 12 and 11 terms for the S-XRD and NPD models, respectively) and 9 additional parameters describing three Gaussian features in the S-XRD patterns from the cryostat; 2 parameters for zero-point error and scale; 3 isotropic displacement parameters (with each site constrained by element); individual atomic displacement modes (with the $C222_1$, $Cmc2_1$, and $P2_1$ models containing 21, 27, and 45 modes, respectively); 2 strain modes transforming as the $\Gamma_1^+$ irrep and 1 or 2 FA modes transforming as $\Gamma_5^+$, depending on the order parameter direction (1 for $C222_1$, $\Gamma_5^+$ (a,√3a), and $Cmc2_1$, $\Gamma_5^+$ (a,0), while 2 for $P2_1$, $\Gamma_5^+$ (a,b)); additional anisotropic broadening terms (7 in the orthorhombic models and 10 in the monoclinic model); and two crystallite size broadening terms were used in the S-XRD refinements. In each phase, the following displacement modes were refined: in $C222_1$, $\Gamma_1^+$ (a), $\Gamma_5^+$ (a,√3a), $\Gamma_1^-$ (a), and $\Gamma_5^-$ (a,0); in $Cmc2_1$, $\Gamma_1^+$ (a), $\Gamma_5^+$ (a,0), $\Gamma_1^-$ (a), $\Gamma_2^-$ (a), and $\Gamma_5^-$ (0,a); and in $P2_1$, $\Gamma_1^+$ (a), $\Gamma_2^+$ (a), $\Gamma_5^+$ (a,b), $\Gamma_1^-$ (a), $\Gamma_2^-$ (a), and $\Gamma_5^-$ (a,b). A small (~0.6 wt.%) impurity phase of 3C-BaTiO$_3$ was visible only in the S-XRD data, which we included in our subsequent fits. Residual deficiencies in the S-XRD Rietveld fits stem largely from domain-wall-like scattering between twin-related diffraction peaks – these could not be accounted for using additional peak broadening parameters. We detail our approach to model these features in Supplementary Note 9.

Table S1 shows that at 100 K and 150 K ($T_c < T < T_o$), the $C222_1$ model provides the most satisfactory fit against the S-XRD data due to the balance between fit quality and the minimal number of refined parameters. Similarly, the $Cmc2_1$ model provides the superior fit against the S-XRD data at both 10 K and 40 K ($T < T_c$) though our ability to resolve a monoclinic peak splitting at 40 K (Fig. 2c) shows a $P2_1$ model is more appropriate at this temperature. This is substantiated by our strain-mode analysis in Supplementary Note 6. The S-XRD data are highly sensitive to the precise symmetry assignment due to the exceptional peak resolution obtained using a multi-analyzer crystal ($\Delta d/d \approx 10^{-4}$ vs $\Delta d/d \approx 10^{-3}$ for the NPD data), giving greater confidence in the symmetry assignments proposed in the main text.

Table S1. Summary of fitting statistics for single-phase Rietveld fits performed against synchrotron XRD (S-XRD) and neutron powder diffraction (NPD) data at select temperatures. The number of fitting parameters ($P$) for each model has also been specified for reference. Note the lowest temperatures in the S-XRD and NPD experiments were different.

| T (K) | Model | S-XRD $R_{wp}$ (%) | $P$(S-XRD) | NPD $R_{wp}$ (%) | $P$(NPD) |
|---|---|---|---|---|---|
| 150 | $C222_1$ | 7.89 | 60 | 6.19 | 47 |
|  | $Cmc2_1$ | 15.2 | 66 | 6.19 | 53 |
|  | $P2_1$ | 6.81 | 88 | 6.17 | 75 |
| 100 | $C222_1$ | 12.8 | 60 | 5.61 | 47 |
|  | $Cmc2_1$ | 24.2 | 66 | 5.93 | 53 |
|  | $P2_1$ | 12.3 | 88 | 5.57 | 75 |
| 40 | $C222_1$ | 25.8 | 60 | 5.18 | 47 |
|  | $Cmc2_1$ | 9.77 | 66 | 4.88 | 53 |
|  | $P2_1$ | 8.87 | 88 | 4.92 | 75 |
| 10 (S-XRD), 20 (NPD) | $C222_1$ | 26.0 | 60 | 5.34 | 47 |
|  | $Cmc2_1$ | 10.5 | 66 | 4.99 | 53 |
|  | $P2_1$ | 9.12 | 88 | 5.05 | 75 |

Illustrative lattice parameters obtained from single-phase symmetry-mode Rietveld fits against the NPD data are shown in Fig. S4. For this, the structural model used only a single phase which was parameterized in terms of $P2_1$ symmetry throughout the entire temperature range to enable $\varphi(\Gamma_5^+)$ to be tracked concomitantly with the lattice parameters. We opted to use the neutron data for this purpose as the diffuse nature of some of the scattering features in the S-XRD data prevented us from obtaining a reliable estimation of the lattice parameters for the domain-wall-like regions using a conventional Rietveld model, hence the lattice parameters shown may be regarded as the average variation in lattice parameters across the full microstructure. We find no evidence of any volume discontinuity around $T_c \approx 50$ K, showing the ferroelectric transition does not exhibit the behavior expected for a conventional first-order phase transition. The average changes in $\varphi(\Gamma_5^+)$ are mirrored in the variation of the $b$ and $\gamma$ lattice parameters of the $P2_1$ cell and reflect the crossover between $Cmc2_1$ and $C222_1$ symmetries across the FE transition. $\varphi(\Gamma_5^+)$ shows an unusual non-monotonic variation above $T_c$ which appears to reflect the increasing fluctuations of the $\Gamma_5^-$ OPD. Thus, although we have insufficient resolution to resolve the diffuse scattering contributions directly in our NPD data, subtle signatures of the Goldstone character of the $\Gamma_5^-$ mode still manifest.

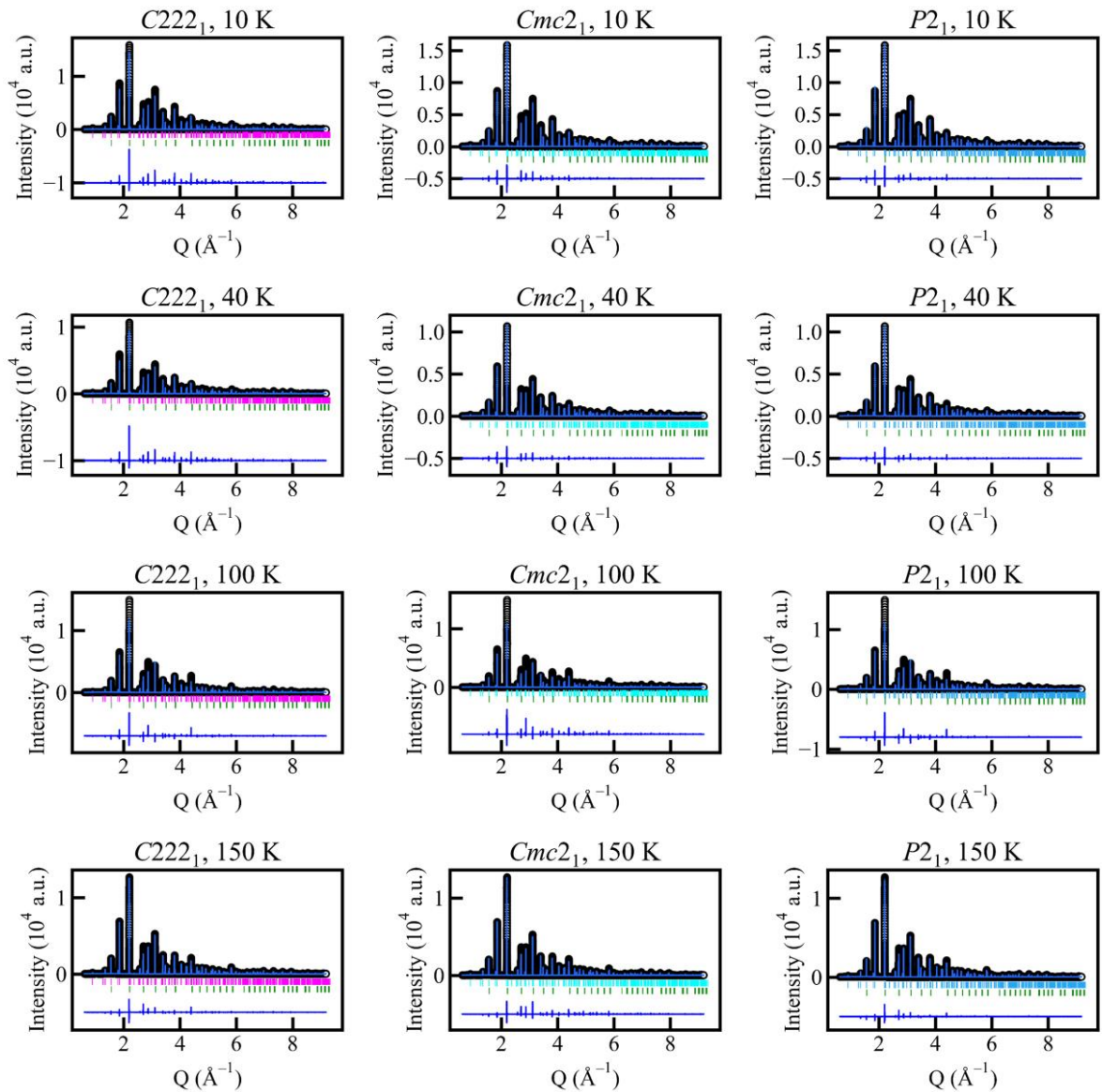

Figure S2. Rietveld fits of the single-phase $C222_1$, $Cmc2_1$, and $P2_1$ structural models against S-XRD data collected at select temperatures below $T_o$. Pink, cyan, and blue tick marks correspond to the expected Bragg reflections of the $C222_1$, $Cmc2_1$, and $P2_1$ symmetries, respectively, while the green tick marks denote reflections from a minor 3C-BaTiO$_3$ impurity (~0.6 wt.%).

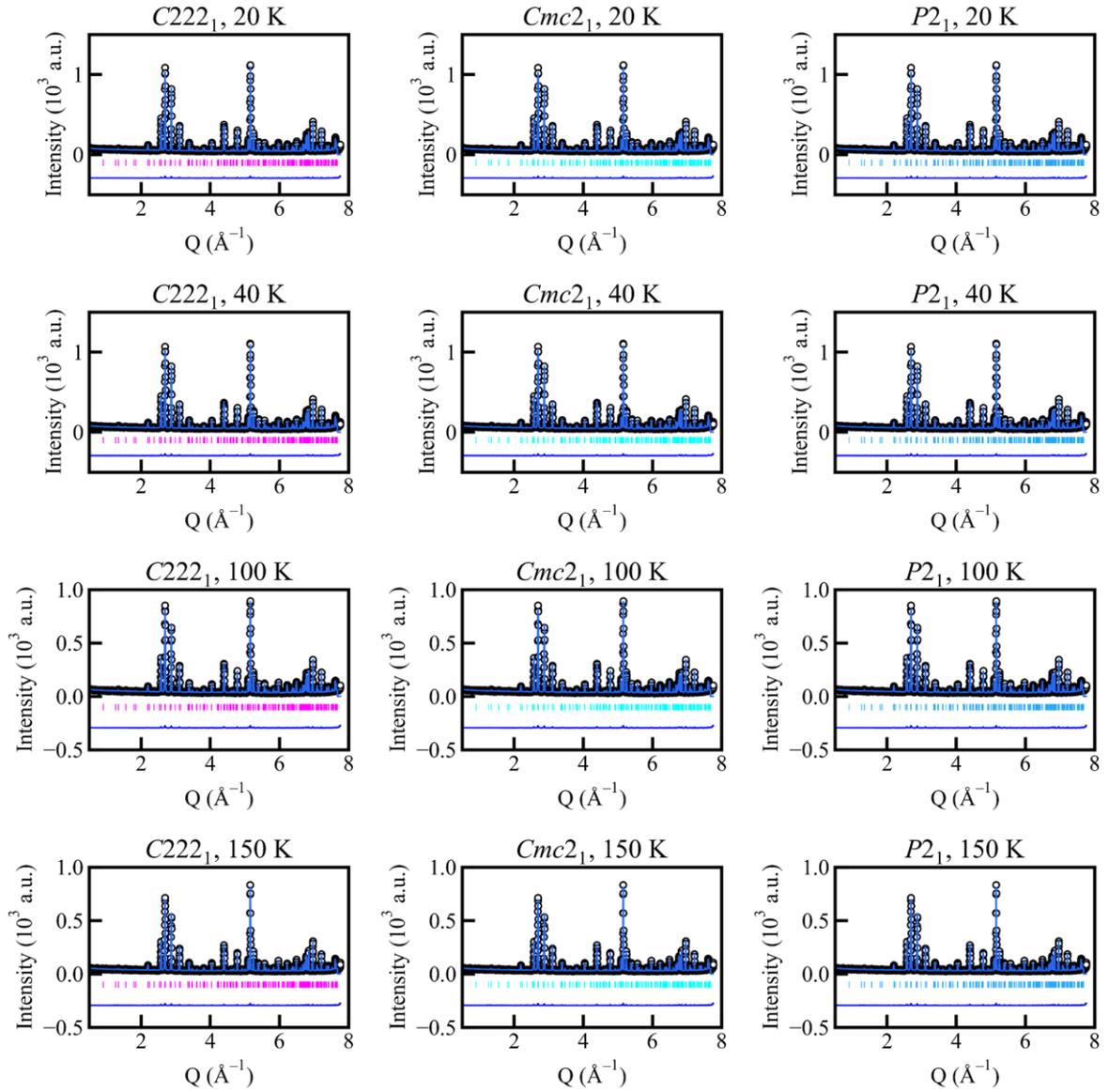

Figure S3. Rietveld fits of the single-phase $C222_1$, $Cmc2_1$, and $P2_1$ structural models against NPD data collected at select temperatures below $T_o$. Pink, cyan, and blue tick marks correspond to the expected Bragg reflections of the $C222_1$, $Cmc2_1$, and $P2_1$ symmetries, respectively. No 3C-$BaTiO_3$ impurity peaks could be distinguished in the NPD patterns.

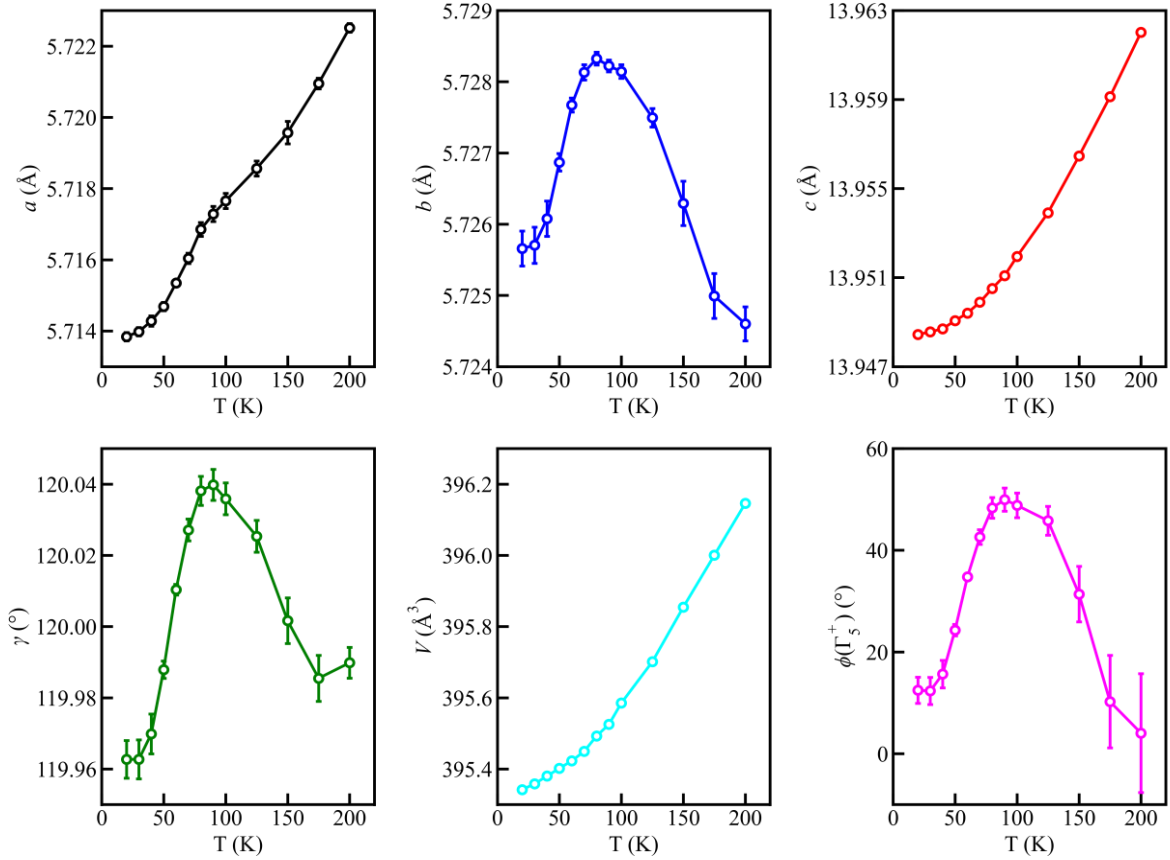

Figure S4. Lattice parameters and fitting statistics as a function of temperature based on a single-phase symmetry-mode Rietveld fit with an unconstrained $P2_1$ model against the NPD data collected below $T_o$. Some error bars are smaller than the data points. Fitting statistics across the entire temperature range are on the order of $R_{wp} \approx 4–6\%$. Note the lower peak resolution compared to the S-XRD data means this model effectively represents an average picture of the low-temperature structure which encompasses the scattering contributions from long-range ordered domains as well as short-range ordered domain-wall-like regions, hence $\varphi(\Gamma_5^-)$ does not relax fully into the high-symmetry $Cmc2_1$ ($\varphi = 0°$) or $C222_1$ ($\varphi = 60°$) directions.

Supplementary Note 4. Mode decomposition of the low-temperature $Cmc2_1$ structure.

To assess the relative contribution of each atom type to the $\Gamma_5^-$ displacements in the $Cmc2_1$ structure, we performed a symmetry-mode Rietveld refinement of this model against the NPD data collected at 20 K. We opted to use the neutron data for this purpose to ensure adequate sensitivity to the oxygen atom displacements in the presence of the heavier Ba and Ti atoms, thus ensuring a fair assessment of the relative displacements of each atom type. Note the unconstrained nature of the refinement means that some of the atomic positions are strongly correlated, so we mainly use this model for comparative purposes. Details of the mode decomposition with respect to the parent $P6_3/mmc$ structure are given in Table S2. The Ti displacements clearly form the largest-amplitude contributions to the total $\Gamma_5^-$ mode, while the Ba and O displacements form smaller contributions. The Ti(1) and Ti(2) $\Gamma_5^-$ displacements are comparable to each other ($d_{max}$(Ti(1)) ≈ $d_{max}$(Ti(2))), showing there is no significant difference in the magnitudes of the SOJT distortions between the two different Ti environments. The $d_{max}$ values are also comparable to that observed in 3C-BaTiO$_3$, where $d_{max}$(Ti) ≈ 0.1 Å in the $R\bar{3}$ phase. Table S3 shows the overall pattern of Ti displacements is analogous to that observed in the TiO$_6$ octahedral environments in 3C-BaTiO$_3$.

Table S2. Mode amplitudes of the antipolar $\Gamma_5^-$ displacements extracted from a single-phase $Cmc2_1$ structural model refined against NPD data collected at 20 K. The amplitudes ($Q$) have been normalized with respect to the $P6_3/mmc$ parent structure obtained from Rietveld fits of the 290 K NPD data (note there is no difference in $Q$ if the normalization is performed with respect to the orthorhombic supercell instead). The relative weights of each mode have been expressed as a percentage of the total mode amplitude ($Q_i^2 / \Sigma Q_i^2$ x 100%), and the maximum displacement observed due to the action of each mode ($d_{max}$) has also been shown for reference.

| Site | Symmetry | $Q(\Gamma_5^-)$ (Å) | Weight (%) | $d_{max}$ (Å) |
|---|---|---|---|---|
| Ba(2) | E(a) | 0.06846 | 4.9 | 0.03423 |
| Ti(1) | E$_u$(a) | -0.17260 | 31.4 | 0.12205 |
| Ti(2) | E(a) | 0.23516 | 58.3 | 0.11758 |
| O(1) | B$_1$(a) | 0.06133 | 4.2 | 0.03634 |
| O(2) | A'$_1$(a) | -0.02592 | 0.7 | 0.01042 |
|  | A'$_2$(a) | 0.00114 | 0.0 | 0.00047 |
|  | A''(a) | -0.01863 | 0.4 | 0.00702 |
|  | Total | 0.30751 | 100.0 |  |

Table S2. Select Ti–O bond data for 6H-BaTiO$_3$ based on the freely refined $Cmc2_1$ model against NPD data collected at 20 K. Reference data for the TiO$_6$ octahedra in 3C-BaTiO$_3$ are also given based on the $R\bar{3}$ model reported in Ref. [45]. Each set of Ti–O distances may be grouped into values which lie within 3σ of each other, and this reproduces the same pattern of 3 x short / 3 x long Ti–O distances observed in the $R\bar{3}$ phase of the 3C polytype.

| Bond length | Distance (Å) | Bond length | Distance (Å) |
|---|---|---|---|
| Ti(1)–O(2) | 1.90(2) x 1 | Ti(2_1)–O(1) | 1.88(1) x 2 |
|  | 1.92(1) x 2 |  | 1.93(2) x 1 |
|  | 2.06(1) x 2 | Ti(2_1)-O(2) | 2.06(2) x 1 |
|  | 2.09(2) x 1 |  | 2.10(1) x 2 |
| Ti–O (3C-BaTiO$_3$) [45] | 1.88 x 3 | Ti(2_2)–O(1) | 1.87(2) x 1 |
|  |  |  | 1.91(1) x 2 |
|  | 2.14 x 3 | Ti(2_2)-O(2) | 2.02(1) x 2 |
|  |  |  | 2.08(2) x 1 |

Supplementary Note 5. Details of the geometry-relaxed structures obtained by DFT calculations.

Table S4. Total energies and mode details of DFT-relaxed structures of 6H-BaTiO$_3$. Relative energies ($\Delta E$) have been reported relative to the lowest-energy structure. Mode amplitudes have been normalized with respect to the DFT-relaxed $P6_3/mmc$ structure. We use the $C222_1$ structure to estimate the stabilization energy of the <120>$_h$-type Ti displacements as 1.235 meV/f.u. per octahedral face. Based on the $P6_3mc$ structure, the $\Gamma_2^-$ <001>$_h$ displacive components offer a comparative stabilization energy of 1.108 meV/f.u. per face, hence the SOJT distortions prefer to orient with respect to the hexagonal basal plane. $\varphi(\Gamma_5^-)$ in the DFT-relaxed $P2_1$ structure is 88.6°, which is very close to the expected value for one of the $Cmc2_1$ domains ($\varphi = 90°$).

| Phase | $P6_3/mmc$ | $P6_3mc$ | $C222_1$ | $Cmc2_1$ | $P2_1$ |
|---|---|---|---|---|---|
| $\Delta E$ (meV/f.u.) | 7.590 | 5.375 | 0.181 | 0.0095 | 0.00 |
| Primary irrep | N/A | $\Gamma_2^-$ (a): 0.26564 | $\Gamma_5^-$ (a,0): 0.34977 | $\Gamma_5^-$ (a,$\sqrt{3}$a/3): 0.35129 | $\Gamma_5^-$ (a,b):<br>a = 0.00503<br>b = 0.20110<br>Total = 0.35072 |
| Secondary irreps | N/A | $\Gamma_1^+$ (a): 0.01479 | $\Gamma_1^+$ (a): 0.01384<br>$\Gamma_5^+$ (a,$\sqrt{3}$a): 0.06504<br>$\Gamma_1^-$ (a): 0.01023 | $\Gamma_1^+$ (a): 0.01441<br>$\Gamma_5^+$ (a,0): 0.03103<br>$\Gamma_2^-$ (a): 0.05330 | $\Gamma_1^+$ (a): 0.01437<br>$\Gamma_2^+$ (a): 0.00005<br>$\Gamma_5^+$ (a,b): 0.03101<br>$\Gamma_1^-$ (a): 0.00077<br>$\Gamma_2^-$ (a): 0.05405 |
| Strain modes | N/A | N/A | $\Gamma_5^+$ (a,$\sqrt{3}$a): –0.00357 | $\Gamma_5^+$ (a,0): –0.00265 | $\Gamma_5^+$ (a,b):<br>a = 0.00115<br>b = 0.00238<br>Total = 0.00264 |

Table S5. Comparison of the total displacive mode amplitudes ($Q$) obtained in the DFT-relaxed $Cmc2_1$ structure to the experimental $Cmc2_1$ structural models based on symmetry-mode refinements against the NPD and S-XRD data at 20 K and 10 K, respectively. The mode amplitudes were normalized with respect to the experimental $P6_3/mmc$ structure obtained from Rietveld fits against NPD data at 290 K. We find that DFT underestimates the amplitude of the polar mode compared to the experimentally obtained mode amplitudes, but otherwise there is good agreement between theory and experiment.

| Mode | DFT $Q$ (Å) | NPD $Q_{20\,K}$ (Å) | S-XRD $Q_{10\,K}$ (Å) |
|---|---|---|---|
| $\Gamma_5^-$ | 0.35243 | 0.30751 | 0.29084 |
| $\Gamma_1^+$ | 0.03106 | 0.03535 | 0.04582 |
| $\Gamma_5^+$ | 0.03113 | 0.05267 | 0.03188 |
| $\Gamma_2^-$ | 0.05351 | 0.13565 | 0.12870 |

Supplementary Note 6. Strain-mode refinements.

Our single-phase symmetry-mode refinements against S-XRD and NPD data showed that the $Cmc2_1$ structural model gave similar qualities of fit to the $P2_1$ model. However, since these two symmetry assignments involve only Γ-point distortions from the $P6_3/mmc$ structure, we are limited in sensitivity to the subtle atomic displacements which distinguish these symmetries. This means the atomic positions will be highly correlated when modelled by Rietveld refinement alone. To overcome this problem, we exploit an alternative approach which relies on the intrinsic coupling between the $\Gamma_5^-$ mode and the ferroelastic strain component transforming as the $\Gamma_5^+$ irrep, as outlined in the main text. The motivation for parameterizing our refinements in terms of the strain mode is the exceptional peak resolution offered by synchrotron XRD measurements ($\Delta d/d \approx 10^{-4}$) – this offers us particular sensitivity to precise peak positions, and as the $Cmc2_1$ and $C222_1$ structures are distinguished by discrete OPDs of the $\Gamma_5^+$ strain mode, this offers a more reliable way to determine the overall symmetry of the system.

To probe the changes in average symmetry across the FE transition, we performed a series of Rietveld fits using fixed variations of the $\Gamma_5^+$ phase angle ($\varphi = \tan^{-1}(b/a)$, where a and b are the components of the $\Gamma_5^+$ OPD). Different values of $\varphi(\Gamma_5^+)$ correspond to distinct $\Gamma_5^+$ OPDs which distinguish the $C222_1$ and $Cmc2_1$ symmetries according to the order parameter space shown in Fig. 1g. We constructed a series of structural models based on a linear interpolation between the expected strain states of the $C222_1$ ($\varphi = 60°$) and $Cmc2_1$ ($\varphi = 0°$) symmetries to assess how the global symmetry changes upon cooling below $T_c$. To perform these refinements, we first obtained a good starting $Cmc2_1$ model with a fixed $\varphi(\Gamma_5^+) = 0°$ against the 10 K dataset. Following this, we fixed the internal structural variables (i.e., the atomic displacement parameters and the refined eigenvalues of the atomic displacement modes) as well as the peak broadening terms, thus reducing the number of fitting parameters in our models. We then constructed a series of structural models in which $\varphi(\Gamma_5^+)$ was incremented in fixed 5° steps to the $C222_1$ limit of $\varphi(\Gamma_5^+) = 60°$. The only refined parameters at each $\varphi(\Gamma_5^+)$ increment were the background terms, the zero-shift and scale parameters, the $\Gamma_1^+$ strain modes (to account for thermal expansion), and the amplitude of the $\Gamma_5^+$ strain $\rho(\Gamma_5^+)$. This series of refinements was then repeated at select temperatures above and below $T_c$ to determine the optimum $\varphi(\Gamma_5^+)$ angle across the FE transition. Fig. 2d in the main text illustrates the results of these fits.

At 10 K, $R_{wp}$ is minimized for $0° \leq \varphi(\Gamma_5^+) \leq 10°$ demonstrating the system tends towards the $Cmc2_1$ limit upon cooling below the FE transition. At 50 K, the best-fitting models are obtained using $20° \leq \varphi \leq 25°$, which is consistent with the average symmetry being best described by the intermediate $P2_1$ structure at this temperature. At 60 K, the best-fitting models are obtained using $40° \leq \varphi \leq 50°$, showing the system still exhibits $P2_1$ symmetry on average but now towards the $C222_1$ limit instead. At 100 K, the best-fitting models are obtained using $50° \leq \varphi \leq 60°$ showing the system has largely transitioned into the $C222_1$ state by this temperature.

Supplementary Note 7. Electron microscopy and electron diffraction experiments.

Electron microscopy and diffraction experiments were carried out in a Jeol 2100 transmission electron microscope operated at 200 kV. A Gatan Model 636 liquid nitrogen cooling holder was used to bring the sample from room temperature to 100 K. The electron diffraction results presented in Figure S5(a)-(d) show that extra spots emerge in the [1$\bar{1}$00] direction with respect to the $P6_3/mmc$ aristotype diffraction pattern upon cooling below $T_o$, clearly indicating a phase transition. Close to $T_c$, these extra spots decrease in intensity as the material approaches the transition into the $Cmc2_1$ phase. However, in the bright field images in Figure S58(e)-(f), no clear contrast indicative of the domain structure is apparent at the lower temperature. A study conducted in the [1000] aristotype zone axis yielded the same results. The material therefore appears homogeneous at the length scales accessible to transmission electron microscopy.

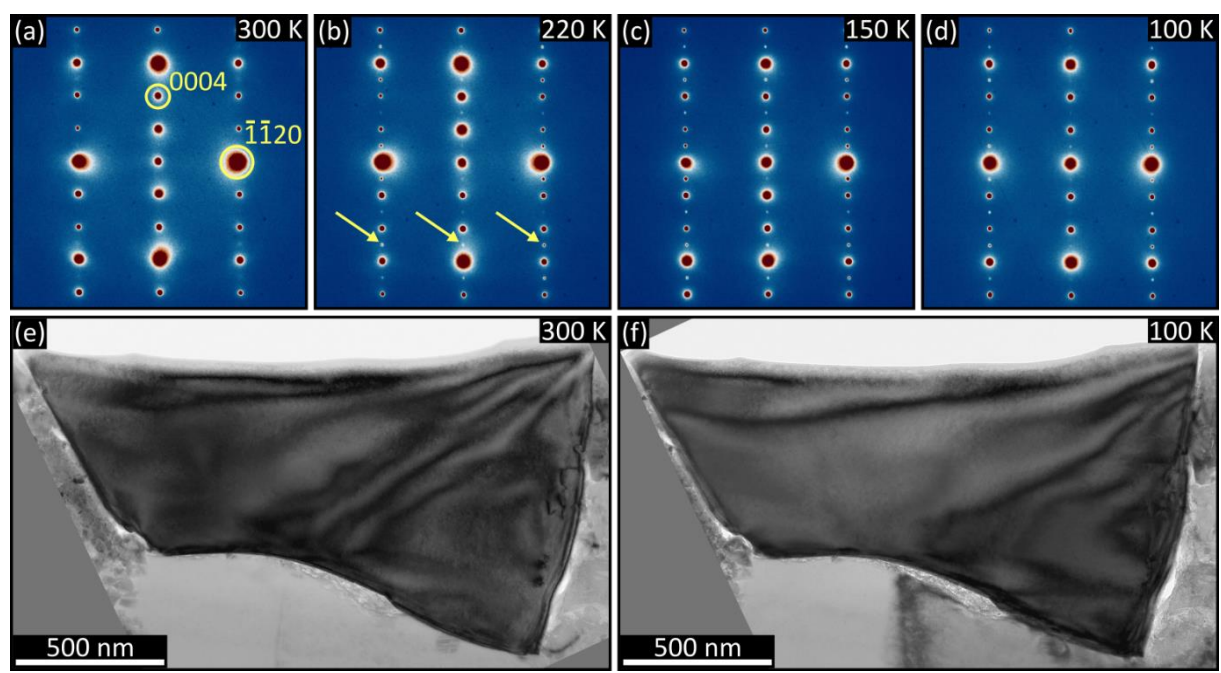

Figure S5. Electron diffraction patterns in the [1$\bar{1}$00] zone axis of the $P6_3/mmc$ aristotype structure recorded at 300 K (a), 220 K (b), 150 K (c) and 100 K (d). The arrows in (b) points to some of the extra spots present in the $C222_1$ and $P2_1$ phases. Bright field micrographs recorded along the same zone axis as the diffraction patterns at 300 K (e) and 100 K (f) show that no domain structure is evident at the lower temperature.

Supplementary Note 8. Details of the 3D-XRD measurements and the strain reconstruction.

The scanning 3D-XRD data were processed using the ImageD11 software package [46]. The sinogram data contains 681 x 3620 diffraction frames corresponding to the translation of the diffractometer across the X-ray beam and the rotation steps. The raw experimental data are available at doi.org/10.15151/ESRF-ES-1734554406. These frames were segmented to identify diffraction spots and these were corrected [47] and converted to reciprocal space co-ordinates using the instrument calibration from a silicon single crystal. The peaks were indexed using an averaged hexagonal unit cell with a = 5.7198 Å and c = 13.9468 Å and hkl indices were assigned to each peak.

Reconstruction of the crystal shape proceeded smoothly using tomographic back projection of the peak intensities versus diffractometer y co-ordinates. The quality of the shape reconstruction was enhanced using the MLEM algorithm [48] in comparison to a simple filtered back projection. Note that the grain shape is nonconvex with some internal holes (Figure S6). Local variations in the unit cell parameters within the crystal were fitted using an adaptation of the algorithm recently described by Henningsson and Hall [49] and using the ASTRA package [50]. Within each pixel we used the usual computation of powder diffraction peak positions via a metric tensor:

$$1/d^2 = A\,h^2 + B\,k^2 + C\,l^2 + D\,kl + E\,hl + F\,hk$$

A uniform image of each of these parameters (*A–F*) was then forward projected to compute a sinogram image by weighting each of the six layers according to the grain shape reconstruction. The value of $1/d^2$ was then computed for each hkl projection using these sinograms of the metric tensor elements. These were compared to the experimental peak positions and a fit was optimized using the L-BFGS algorithm within the scipy package. Fortunately, the optimization problem is linear. No symmetry constraints or regularization were applied and this leads to some high frequency noise in components that should be zero. In comparison to the usual methods for local strain refinement, this metric tensor-based approach is free of any selection of a reference "*d*-zero" unit cell. Instead, we refined the local lattice parameters within each voxel. The resulting variation in lattice parameters is shown in Figure S7.

Unit cells were computed for each pixel from the metric tensor elements. The local strain in each voxel was parameterized in terms of the $\Gamma_5^+$ strain components symmetrized with respect to the room-temperature *P6₃/mmc* aristotype structure. To do this, we expressed the refined lattice basis vectors from each voxel in cartesian coordinates and compared these to the basis vectors of the parent cell to obtain unitless parent cell strains $P_1$, $P_2$, and $P_6$ according to the following relationships:

$$P_1 = \frac{a(x) - a_0(x)}{a_0(x)}$$

$$P_6 = \frac{2(b(x) - b_0(x)*(P_1 + 1)}{b_0(y)}$$

$$P_2 = \frac{b(y) - b_0(x) * (P_6/2)}{b_0(y)} - 1$$

where a(x), b(x), and b(y) refer to cartesian components of the distorted lattice basis vectors, and $a_o(x)$, $b_o(x)$, and $b_o(y)$ refer to those of the parent lattice. From these, we computed the $\Gamma_5^+$ (a,b) OPD components according to $\Gamma_5^+$ (a,0) = $(P_1 - P_2)/\sqrt{2}$ and $\Gamma_5^+$ (0,a) = $P_6/\sqrt{2}$. By treating the $\Gamma_5^+$ strain as a two-dimensional order parameter ($\rho$, $\varphi$), we mapped the variation in $\varphi(\Gamma_5^+)$ across the image of the crystal, as shown in the main text. The raw $\Gamma_5^+$ OPD components are shown in Figure S8, while $\varphi(\Gamma_5^+)$ and $\rho(\Gamma_5^+)$ are shown in Figure S9.

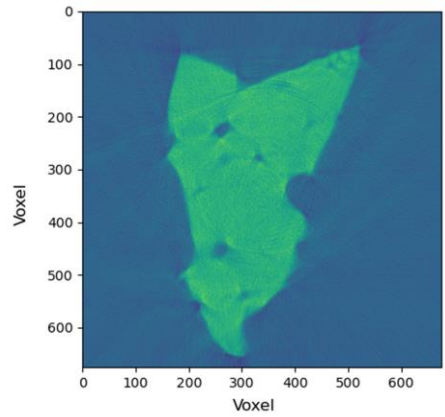

Figure S6. Reconstruction of the crystal shape based on a tomographic back projection of the peak intensities versus the diffractometer y co-ordinates.

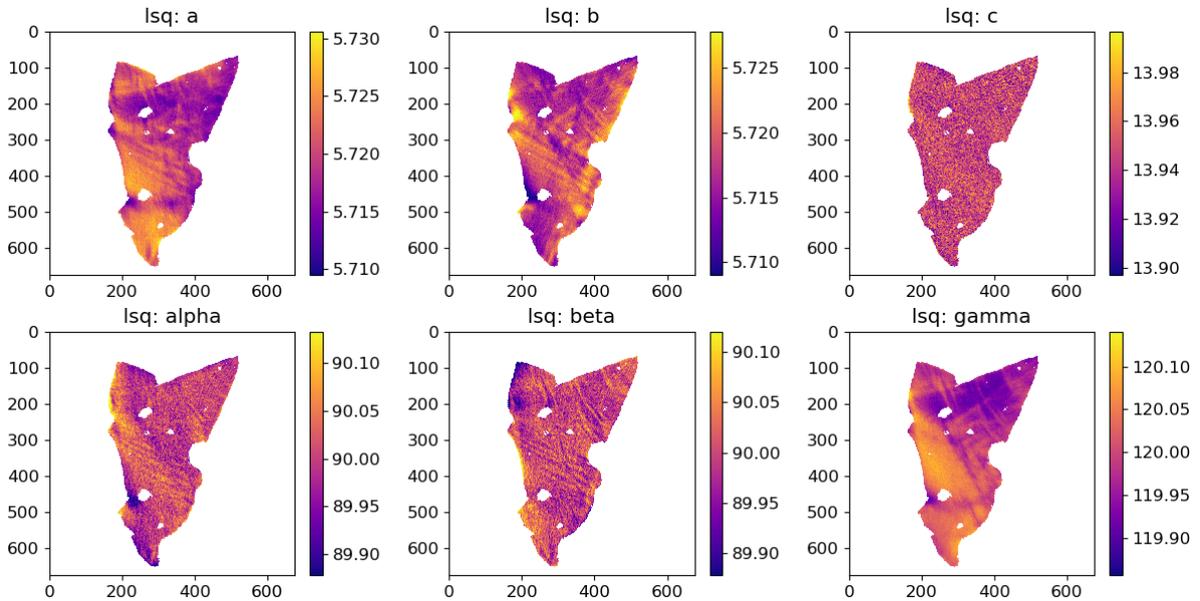

Figure S7. Local variations in lattice parameters obtained from least-squares (lsq) fits against our 3D-XRD data. Each plot is labelled with the corresponding lattice parameter. The units on the x- and y-axes correspond to the voxels of the image.

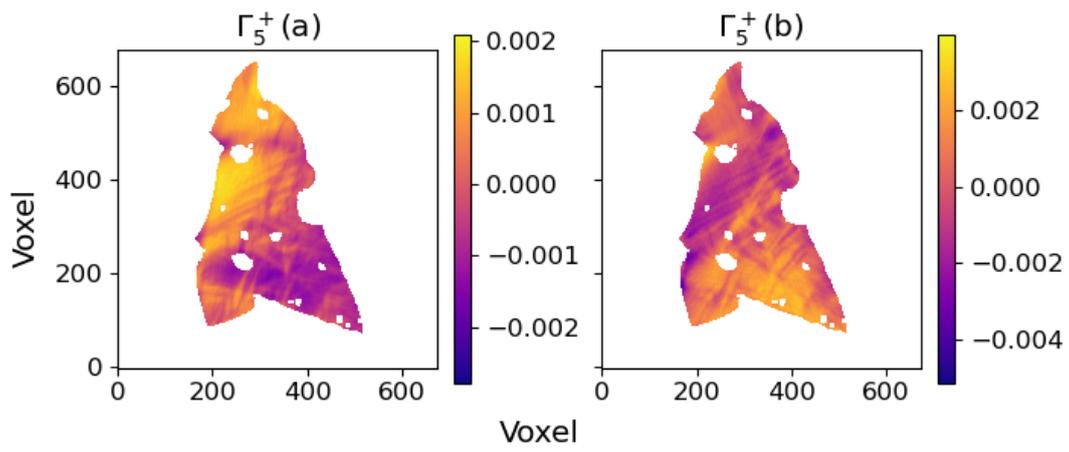

Figure S8. Map of the variation in the $\Gamma_5^+$ OPD components across the crystal.

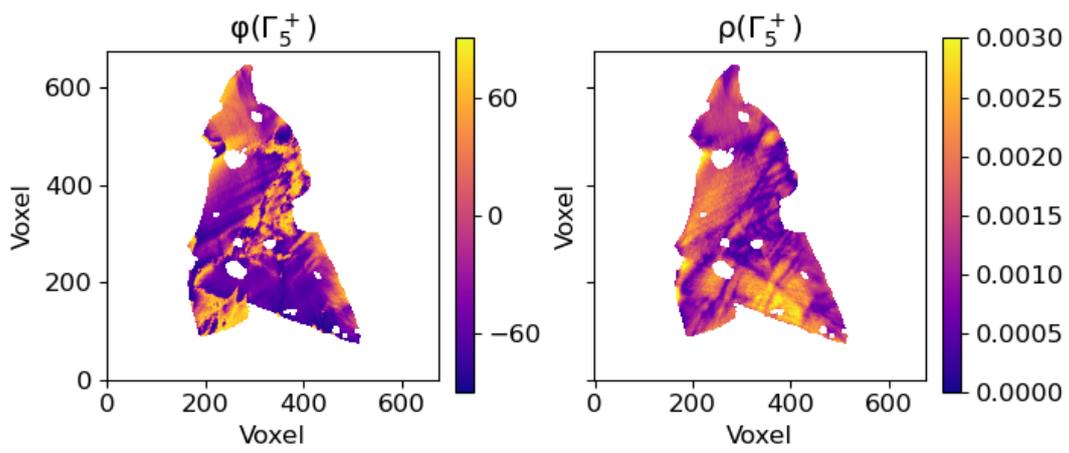

Figure S9. Map of the variations in $\varphi(\Gamma_5^+)$ and $\rho(\Gamma_5^+)$ across the crystal.

Supplementary Note 9. Domain-wall scattering contributions.

Close inspection of our Rietveld fits against S-XRD data revealed additional broadening between twin-related diffraction peaks (Fig. S10a). These could not be accounted for by using anisotropic strain or size broadening terms in our structural models. Such inter-peak broadening is characteristic of coherent scattering contributions from domain walls. Previous approaches to model such diffuse-like scattering contributions have utilized a secondary phase component with either anisotropic atomic displacement parameters, a lower crystal symmetry, or additional anisotropic size/strain broadening terms. However, such approaches inevitably introduce an excessive number of refined parameters into the total structural model, limiting its meaningfulness. We also found that this approach did not give a satisfactory account of the domain-wall-like scattering in our diffraction patterns (Fig. S10b). For these reasons, we sought an alternative method to describe the domain-wall scattering contributions.

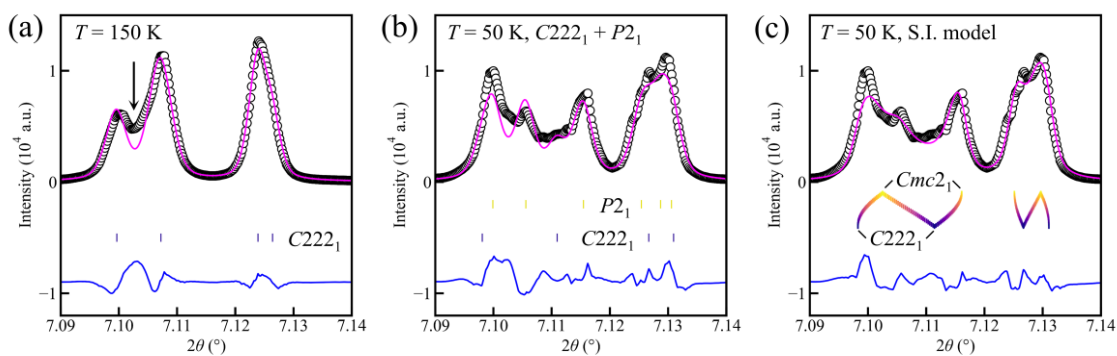

Figure S10. Modelling domain-wall-like scattering contributions in 6H-BaTiO$_3$. (a) Excerpt of a single-phase $C222_1$ Rietveld fit against S-XRD data collected at 150 K. The arrow highlights the domain-wall-like scattering which is not described by the single-phase model. (b) Excerpt of a two-phase Rietveld model consisting of $P2_1$ and $C222_1$ phases against the data collected at 50 K. Here, the two-phase model captures some of the domain-wall-like scattering, but ultimately it fails to capture the complex mixture of diffuse and Bragg scattering contributions near the FE transition. (c) Excerpt of a representative strain-interpolated (S.I.) model used to obtain the histograms shown in Fig. 4. The fit shown corresponds to the 50 K dataset.

First, we performed a series of two-phase Rietveld refinements where a second phase was introduced featuring a distinct ferroelastic strain to account for the differing peak positions of the diffuse-like scattering contributions in the diffraction patterns. We constrained the internal structural details and peak broadening terms for the two phases to be equal to each other to limit the number of refined parameters in our models; given the limiting $C222_1$ and $Cmc2_1$ structures should predominantly differ in the relative orientations of their $\Gamma_5^-$ displacements – which produces only minor intensity differences to the existing Bragg reflections – this approach is justified. The internal structural details used were based on a $P2_1$ structure refined against the data collected at 10 K. Both $\rho(\Gamma_5^+)$ and $\varphi(\Gamma_5^+)$ could then be refined for each phase component so that any subsequent improvement in the quality of the fit could be attributed to these fitting parameters. Overall, the structural models contained the usual background terms, a scale factor for each phase component, a single scaling parameter to describe the global temperature variation of the atomic displacement parameters, the totally symmetric $\Gamma_1^+$ strain modes to account for thermal expansion, and $\rho(\Gamma_5^+)$ and $\varphi(\Gamma_5^+)$ in each phase.

We constructed a total of eight structural models based on different possible strain scenarios. The corresponding $\rho(\Gamma_5^+)$ and $\varphi(\Gamma_5^+)$ parameters of these models are depicted with respect to the $\Gamma_5^+$ order parameter space in Fig. S11a, and details of the fitted parameters are shown in Fig. S11b. We found that a model containing two $P2_1$ phases with freely refined $\rho(\Gamma_5^+)$ and $\varphi(\Gamma_5^+)$ (Model 8) gave the best fit against the observed data, which is unsurprising given that it has two additional refined parameters compared to the majority of the other models. Based on this model, $\rho(\Gamma_5^+)$ increases monotonically with respect to the Bragg scattering contributions ($\rho_1$) upon cooling below the FE transition, while there is a clear cusp in $\rho_2(\Gamma_5^+)$ below the FE transition. Model 8 shows $\varphi_1(\Gamma_5^+)$ exhibits a crossover between $C222_1$ and $Cmc2_1$ symmetries across $T_c$, as was captured in our single-phase refinements, but the variation in $\varphi_2(\Gamma_5^+)$ suggests the crossover occurs via an interconversion between the long-range ordered domains and the domain-wall-like regions across the microstructure.

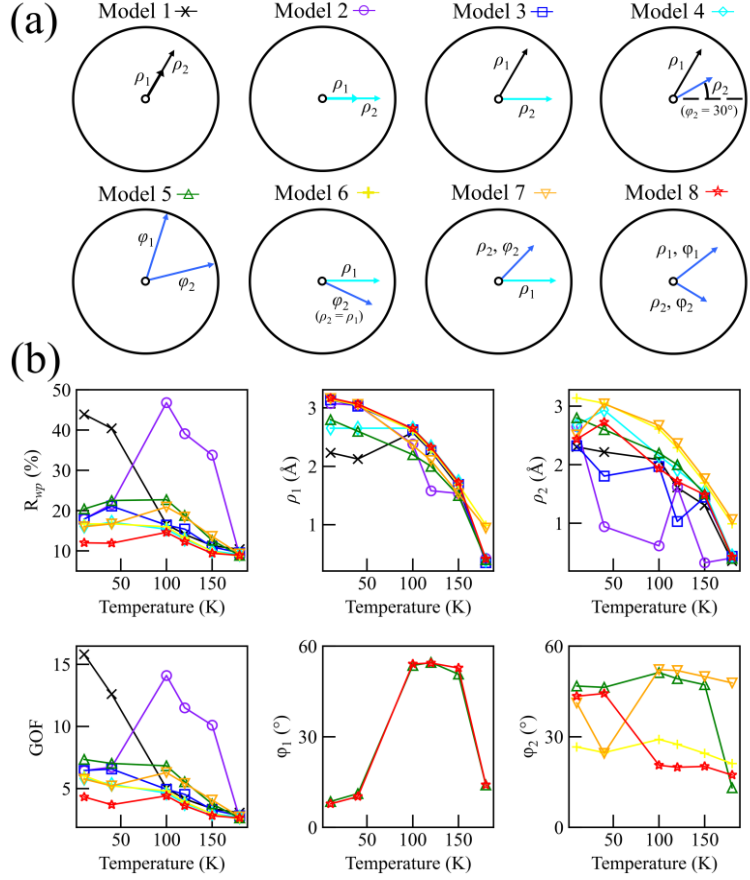

Figure S11. Summary of the two-phase strain mode refinements. (a) The eight strain models tested, depicted with respect to the order parameter space spanned by the $\Gamma_5^+$ (a,b) irrep. The models correspond to the following: Model 1 = two $C222_1$ phases with $\rho(\Gamma_5^+)$ refined separately in both; Model 2 = two $Cmc2_1$ phases, with $\rho(\Gamma_5^+)$ refined separately in both; Model 3 = one $C222_1$ phase and one $Cmc2_1$ phase, with $\rho(\Gamma_5^+)$ refined separately in both; Model 4 = one $C222_1$ phase and one $P2_1$ phase, with $\rho(\Gamma_5^+)$ refined separately in both; Model 5 = two $P2_1$ phases, with $\varphi(\Gamma_5^+)$ refined separately in both; Model 6 = one $Cmc2_1$ phase and one $P2_1$ phase, with $\rho(\Gamma_5^+)$ constrained to be equal in each phase and $\varphi(\Gamma_5^+)$ refined in the $P2_1$ phase; Model 7 = one $C222_1$ phase with $\rho(\Gamma_5^+)$ refined and one $P2_1$ phase with both $\rho(\Gamma_5^+)$ and $\varphi(\Gamma_5^+)$ refined; Model 8 = two $P2_1$ phases with both $\rho(\Gamma_5^+)$ and $\varphi(\Gamma_5^+)$ refined. Note in Models 4, 6, and 7 where an alternate choice of symmetry is possible for the orthorhombic phase, both $C222_1$ and $Cmc2_1$ options were tested so that only the results from the best-fitting model overall are depicted. The first and second phases account for the Bragg and diffuse-like scattering contributions, respectively. (b) Refinement statistics and $\varphi/\rho$ parameters extracted from the eight strain models performed with respect to select temperatures above and below $T_c$. Error bars are smaller than the data points. Each model number has been labelled according to the legend shown in (a).

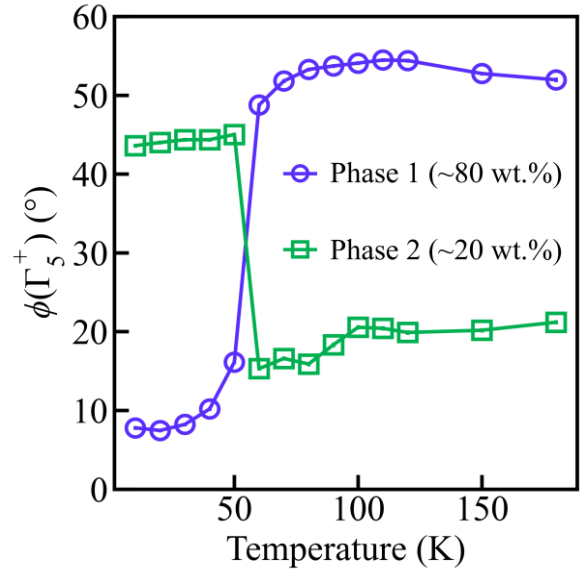

Figure S12. Full temperature variation of $\varphi(\Gamma_5^+)$ based on the Model 8 strain-mode refinement. Purple circles and green squares correspond to the majority phase ("Phase 1") and the domain-wall-like scattering contributions ("Phase 2"), respectively. Error bars are smaller than the data points.

Supplementary Note 10. Strain-interpolated models.

To model the distribution of ferroelastic strains in 6H-BaTiO$_3$, we used the results of our previous two-phase strain models to interpolate between the limiting $C222_1$ and $Cmc2_1$ strain states at various temperatures. To do this, we took the two fitted $\rho(\Gamma_5^+)$ parameters from Model 3 (see Supplementary Note 9) and used these as the range over which to perform the interpolation. A linear interpolation between these values was generated based on 2° increments of $\varphi(\Gamma_5^+)$ in the range 0–60°; given $\rho(\Gamma_5^+)$ varies as a function of temperature (Fig. S11b), this interpolation procedure was repeated at each temperature of interest. Details of the $\rho(\Gamma_5^+)$ range obtained at each temperature are shown in Table S6.

Once the interpolation range was defined for each temperature, 31 structural phases were generated corresponding to the 2° bins of $\varphi(\Gamma_5^+)$ between the $Cmc2_1$ and $C222_1$ limiting strain states. Each "phase" in this model physically corresponds to regions of the sample which possess a particular deviation in $\varphi(\Gamma_5^+)$ away from the high-symmetry OPDs, thus the combination of refined scale factors for each "phase" produces a model of the distribution of strain across the microstructure of the sample. We use an arbitrary bin size of 2° to obtain these distributions, though strictly speaking we lack sufficient resolution in our diffraction experiments to distinguish such small variations in $\varphi(\Gamma_5^+)$. Nevertheless, we found that using larger bin sizes such as 4° and 8° did not change the overall shape of the distribution we obtained, nor did it give any improvement in the final fitting statistics.

As with our previous strain models, the internal structural details of each phase were kept fixed to minimize the number of refined parameters. Each strain model consisted of the usual 21 background terms from our previous Rietveld refinements, a global scaling factor for the isotropic atomic displacement parameters, 2 symmetric $\Gamma_1^+$ strain terms accounting for thermal expansion, two global isotropic size broadening terms (constrained to be equal for each phase), and a scale factor for each of the 31 phases (58 parameters in total). We found that the inclusion or omission of the size broadening terms had little effect on the final distribution obtained, but their inclusion did give a better fit to the observed peak shapes (e.g., for the 10 K dataset $R_{wp}$ = 10.4% with the broadening refined while $R_{wp}$ = 21.0% without these terms included). To prevent biasing the distributions in subsequent models, the scales for each phase were reset to a fixed starting value before being allowed to refine freely against the experimental data.

To generate the histograms depicted in Fig 4, the refined scale factors for each of the 31 phases were extracted from the fits at each temperature. As powder diffraction is insensitive to the absolute orientations of the different twin domains in the low-temperature structures, the refined scale factors represent an average of the scattering contributions from all symmetry-related domains present within the sample, hence we have treated the distributions as symmetric around the high-symmetry $C222_1$ and $Cmc2_1$ OPDs. By tiling the distribution obtained in the range 0–60° across the full 360° order parameter space, we arrive at the final distributions shown in Fig 4.

Table S6. Details of the interpolation procedure used to obtain the strain distributions depicted in Figs. 4c–e. $\rho_1(\Gamma_5^+)$ and $\rho_2(\Gamma_5^+)$ correspond to the fitted parameters extracted from a two-phase strain model consisting of $C222_1$ and $Cmc2_1$ symmetries (Model 3). The corresponding interpolation range ($\Delta\rho$) and fitting statistics for the corresponding strain distributions have also been provided. Note the strain in each individual phase component was normalized with respect to the Rietveld-refined structure at 10 K, so the $\Delta\rho$ values reported only reflect relative differences in strain. Estimates of the full-widths at half-maxima ($\sigma$) based on Gaussian fits of the distributions (shown in Fig. S13) are provided for reference. For the 150 K dataset (*), a single distribution did not give a good description of the data.

| T (K) | $\rho_1(\Gamma_5^+)$ (a.u.) | $\rho_2(\Gamma_5^+)$ (a.u.) | $\Delta\rho$ (a.u.) | $R_{wp}$ (%) | $\sigma$ (°) |
|---|---|---|---|---|---|
| 10  | 3.15 | 2.30 | 0.85 | 10.4 | 9.49(4) |
| 30  | 3.03 | 1.80 | 1.23 | 9.80 | 13.9(1) |
| 40  | 3.03 | 1.80 | 1.23 | 9.78 | 23.7(1) |
| 50  | 2.95 | 2.85 | 0.10 | 12.3 | 92.9(9) |
| 60  | 2.95 | 2.86 | 0.09 | 12.7 | 42.2(3) |
| 70  | 3.03 | 1.80 | 1.23 | 18.9 | 34.9(2) |
| 100 | 2.60 | 1.95 | 0.65 | 16.0 | 8.86(4) |
| 150 | 1.70 | 1.45 | 0.25 | 11.2 | – * |
| 180 | 0.92 | 0.90 | 0.02 | 12.3 | 84.1(7) |

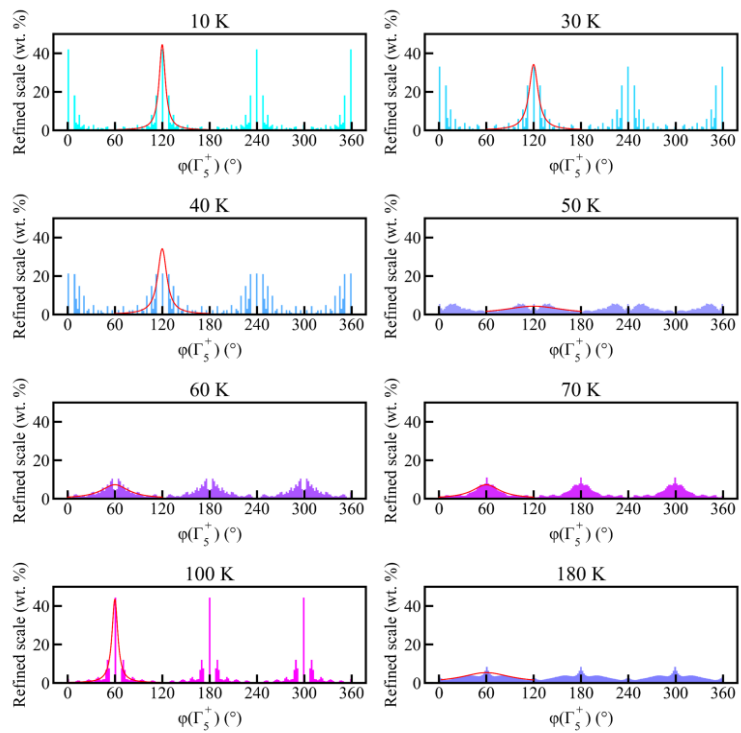

Figure S13. Selected Gaussian fits to the modelled strain distributions at select temperatures, as depicted by the solid red lines. Bins with negligible phase fractions (<0.1 wt.%) were omitted for these fits.

To assess the validity of our strain-interpolated models, we compared the strain distribution obtained from our S-XRD data at 150 K to that obtained from our 3D-XRD data. The distributions are compared in Figure S14. The two approaches produce qualitatively similar distributions at 150 K, particularly in capturing the majority $C222_1$ domains and the overall shape in the distribution formed from minor lower-symmetry inclusions.

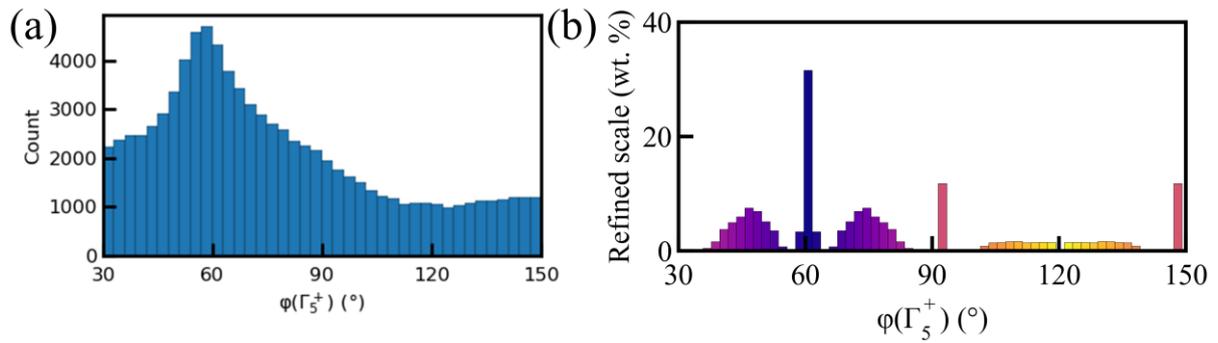

Figure S14. Domain distributions at 150 K based on (a) our reconstructed strain map based on 3D-XRD data and (b) our strain-interpolated model based on high-resolution S-XRD data.

Supplementary Note 11. Preliminary DFT results on the 2H, 4H, 8H, and 10H polytypes.

To investigate whether the SOJT instability could produce Goldstone-like modes in other hexagonal perovskite polytypes, we performed DFT calculations on the 2H, 4H, 8H, and 10H polytypes of the perovskite structure. Representative structures for each polytype were generated in terms of the $BaTiO_3$ composition before performing geometry relaxations in the same manner as for the 6H polytype. For each polytype, the aristotype structure is described by $P6_3/mmc$ symmetry so that the irrep labels are the same as for the 6H polytype and so that they correspond to analogous configurations of atomic displacements. Note $BaTiO_3$ has yet to be stabilized in any of these alternative polytypes, so we merely use these structures to simulate how the SOJT instability manifests in these systems.

Phonon dispersion curves for each geometry-relaxed polytype are shown in Fig. S15. Total energies and symmetry mode decompositions are provided in Tables S7–10. All four polytypes possess unstable $\Gamma$-point modes with either $\Gamma_5^-$ or $\Gamma_2^-$ symmetry. The $\Gamma_5^-$ modes represent the most unstable modes with the exception of the 2H polytype, where we find a $K_3$ mode is the most unstable instead. This is in agreement with phonon calculations on the 2H structure of $BaMnO_3$ [34]. In the case of the 4H, 8H, and 10H polytypes, the $C222_1$, $Cmc2_1$, and $P2_1$ structures possess similar energies relative to the ground state suggesting the energy landscape which distinguishes $C222_1$ and $Cmc2_1$ symmetries is relatively flat. Similarly to the 6H polytype, the relaxed $P2_1$ structures are nearly identical to either of the $Cmc2_1$ or $C222_1$ structures so that the ground states in these polytypes most likely correspond to one of the high-symmetry OPDs of the $\Gamma_5^-$ mode. Thus although we have not computed a full PES for each polytype, the SOJT instability will likely manifest similarly to the picture we have outlined in 6H-$BaTiO_3$, hence its Goldstone character appears to be generalizable to other polytype structures of perovskite materials.

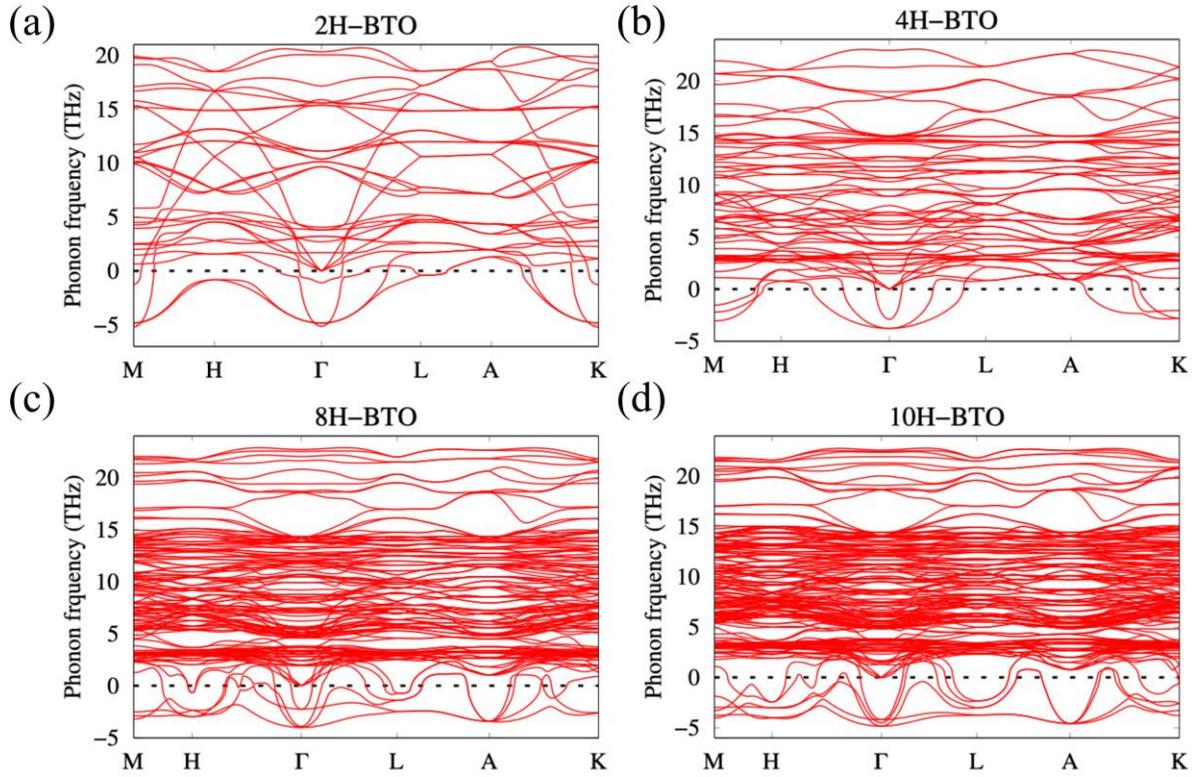

Figure S15. Phonon dispersions of the 2H (a), 4H (b), 8H (c), and 10H (d) polytypes of BaTiO$_3$.

Table S7. Total energies and mode details of DFT-relaxed structures of the hypothetical polytype 2H-BaTiO$_3$. Relative energies (ΔE) have been reported relative to the lowest-energy structure. Mode amplitudes have been normalized with respect to the DFT-relaxed $P6_3/mmc$ structure.

| Phase | $P6_3/mmc$ | $P6_3mc$ | $C222_1$ | $Cmc2_1$ | $P2_1$ |
|---|---|---|---|---|---|
| ΔE (meV/f.u.) | 489.489 | 440.462 | 443.845 | 110.773 | 0.00 |
| Primary irrep | N/A | $\Gamma_2^-$(a): 1.04250 | $\Gamma_5^-$(a,0): 0.30755 | $\Gamma_5^-$(a,√3a/3): 0.91542 | $\Gamma_5^-$(a,b): 1.62147 |
| Secondary irreps | N/A | $\Gamma_1^+$(a): 0 | $\Gamma_1^+$(a): 0.02256<br>$\Gamma_5^+$(a,√3a): 0.07532 | $\Gamma_1^+$(a): 0.21707<br>$\Gamma_5^+$(a,0): 1.18705<br>$\Gamma_2^-$(a): 2.95790 | $\Gamma_1^+$(a): 1.07570<br>$\Gamma_2^+$(a): 2.10833<br>$\Gamma_5^+$(a,b): 1.84872<br>$\Gamma_2^-$(a): 2.71678 |
| Strain modes | N/A | N/A | $\Gamma_5^+$(a,√3a): –0.00436 | $\Gamma_5^+$(a,0): –0.19641 | $\Gamma_5^+$(a,b)<br>a = 0.03644<br>b = 0.21287 |

Table S8. Total energies and mode details of DFT-relaxed structures of the hypothetical polytype 4H-BaTiO$_3$. Relative energies (ΔE) have been reported relative to the lowest-energy structure. Mode amplitudes have been normalized with respect to the DFT-relaxed $P6_3/mmc$ structure.

| Phase | $P6_3/mmc$ | $P6_3mc$ | $C222_1$ | $Cmc2_1$ | $P2_1$ |
|---|---|---|---|---|---|
| ΔE (meV/f.u.) | 13.563 | 11.113 | 0.0146 | 0.1459 | 0.00 |
| Primary irrep | N/A | $\Gamma_2^-$(a): 0.26355 | $\Gamma_5^-$(a,0): 0.32516 | $\Gamma_5^-$(a,√3a/3): 0.32254 | $\Gamma_5^-$(a,b): 0.32569 |
| Secondary irreps | N/A | $\Gamma_1^+$(a): 0.00868 | $\Gamma_1^+$(a): 0.01514<br>$\Gamma_5^+$(a,√3a): 0.07532<br>$\Gamma_1^-$(a): 0.01775 | $\Gamma_1^+$(a): 0.01494<br>$\Gamma_5^+$(a,0): 0.05613<br>$\Gamma_2^-$(a): 0.01078 | $\Gamma_1^+$(a): 0.01510<br>$\Gamma_2^+$(a): 0.00011<br>$\Gamma_5^+$(a,b): 0.06680<br>$\Gamma_1^-$(a): 0.01822<br>$\Gamma_2^-$(a): 0.00262 |
| Strain modes | N/A | N/A | $\Gamma_5^+$(a,√3a): –0.00385 | $\Gamma_5^+$(a,0): –0.00340 | $\Gamma_5^+$(a,b)<br>a = 0.00169<br>b = 0.00346 |

Table S9. Total energies and mode details of DFT-relaxed structures of the hypothetical polytype 8H-BaTiO$_3$. Relative energies (ΔE) have been reported relative to the lowest-energy structure. Mode amplitudes have been normalized with respect to the DFT-relaxed $P6_3/mmc$ structure.

| Phase | $P6_3/mmc$ | $P6_3mc$ | $C222_1$ | $Cmc2_1$ | $P2_1$ |
|---|---|---|---|---|---|
| ΔE (meV/f.u.) | 11.973 | 5.191 | 0.298 | 0.00674 | 0.00 |
| Primary irrep | N/A | $\Gamma_2^-$(a): 0.37862 | $\Gamma_5^-$(a,0): 0.45480 | $\Gamma_5^-$(a,√3a/3): 0.45277 | $\Gamma_5^-$(a,b): 0.45424 |
| Secondary irreps | N/A | $\Gamma_1^+$(a): 0.03150 | $\Gamma_1^+$(a): 0.02632<br>$\Gamma_5^+$(a,√3a): 0.13588<br>$\Gamma_1^-$(a): 0.01656 | $\Gamma_1^+$(a): 0.02768<br>$\Gamma_5^+$(a,0): 0.02441<br>$\Gamma_2^-$(a): 0.10042 | $\Gamma_1^+$(a): 0.02773<br>$\Gamma_2^+$(a): 0.00003<br>$\Gamma_5^+$(a,b): 0.02698<br>$\Gamma_1^-$(a): 0.00074<br>$\Gamma_2^-$(a): 0.09872 |
| Strain modes | N/A | N/A | $\Gamma_5^+$(a,√3a): –0.00499 | $\Gamma_5^+$(a,0): –0.00286 | $\Gamma_5^+$(a,b)<br>a = 0.00126<br>b = 0.00262 |

Table S10. Total energies and mode details of DFT-relaxed structures of the hypothetical polytype 10H-BaTiO$_3$. Relative energies (ΔE) have been reported relative to the lowest-energy structure. Mode amplitudes have been normalized with respect to the DFT-relaxed $P6_3/mmc$ structure.

| Phase | $P6_3/mmc$ | $P6_3mc$ | $C222_1$ | $Cmc2_1$ | $P2_1$ |
|---|---|---|---|---|---|
| ΔE (meV/f.u.) | 16.785 | 5.494 | 0.507 | 0.009836 | 0.00 |
| Primary irrep | N/A | $\Gamma_2^-$(a): 0.53929 | $\Gamma_5^-$(a,0): 0.54085 | $\Gamma_5^-$(a,√3a/3): 0.53696 | $\Gamma_5^-$(a,b): 0.54121 |
| Secondary irreps | N/A | $\Gamma_1^+$(a): 0.03266 | $\Gamma_1^+$(a): 0.03499<br>$\Gamma_5^+$(a,√3a): 0.24672<br>$\Gamma_1^-$(a): 0.02143 | $\Gamma_1^+$(a): 0.03684<br>$\Gamma_5^+$(a,0): 0.03470<br>$\Gamma_2^-$(a): 0.13975 | $\Gamma_1^+$(a): 0.03682<br>$\Gamma_2^+$(a): 0.00011<br>$\Gamma_5^+$(a,b): 0.03447<br>$\Gamma_1^-$(a): 0.00109<br>$\Gamma_2^-$(a): 0.14183 |
| Strain modes | N/A | N/A | $\Gamma_5^+$(a,√3a): –0.00635 | $\Gamma_5^+$(a,0): –0.00329 | $\Gamma_5^+$(a,b)<br>a = 0.00140<br>b = 0.00295 |